\documentclass[prl,twocolumn,superscriptaddress,footinbib]{revtex4-1} 

\usepackage{morefloats}
\usepackage[dvips]{graphicx}
\usepackage{color}
\usepackage{epsfig,graphicx,amsfonts,amsbsy}
\usepackage{amsmath,amsfonts,amsthm,amssymb}
\usepackage{appendix}
\usepackage{makeidx}
\usepackage{url}
\usepackage{verbatim}
\usepackage[bookmarksnumbered,pdfpagelabels=true,plainpages=false,colorlinks=true,linkcolor=blue,citecolor=blue,urlcolor=blue]{hyperref}
\usepackage[rightcaption]{sidecap}
\usepackage{array}
\usepackage{multirow}
\usepackage{tabularx}
\usepackage{braket} 
\usepackage{scrextend}
\usepackage{dsfont}

\newcommand{\td}[1]{\tilde{#1}}
\newcommand{\pt}{\partial}
\newcommand\footnoteref[1]{\protected@xdef\@thefnmark{\ref{#1}}\@footnotemark}

\newcommand{\mb}{\mathbf}
\newcommand{\mc}{\mathcal}
\newcommand{\Jint}{J_{\scalebox{0.6}{int}}}
\newcommand{\mJint}{\mathcal{J}_{\scalebox{0.6}{int}}}

\newcommand{\Phic}{\Phi_{\scalebox{0.6}{cl}}}
\newcommand{\Pic}{\Pi_{\scalebox{0.6}{cl}}}
\newcommand{\Thetac}{\Theta_{\scalebox{0.6}{cl}}}
\newcommand{\Fint}{\mc{F}_{\scalebox{0.6}{int}}}
\newcommand{\Fext}{\mc{F}_{\scalebox{0.6}{ext}}}
\newcommand{\Hext}{H_{\scalebox{0.6}{ext}}}
\newcommand{\Hint}{H_{\scalebox{0.6}{int}}}
\newcommand{\Hrot}{H_{\scalebox{0.6}{rot}}}
\newcommand{\Feff}{\mc{F}_{\scalebox{0.6}{eff}}}
\newcommand{\pext}{\phi_{\scalebox{0.6}{ext}}}
\newcommand{\vp}{\varphi_0}
\newcommand{\hvp}{\hat{\varphi}_0}
\newcommand{\tmn}{\tilde{\mb{n}}}

\begin{document}

\title{Skyrmion Qubits: A New Class of Quantum Logic Elements Based on Nanoscale Magnetization}

\author{Christina Psaroudaki}
\affiliation{Department of Physics and Institute for Quantum Information and Matter, California Institute of Technology, Pasadena, California 91125, USA}
\affiliation{Institute for Theoretical Physics, University of Cologne, D-50937 Cologne, Germany}
\email{cpsaroud@caltech.edu}

\author{Christos Panagopoulos}
\affiliation{Division of Physics and Applied Physics, School of Physical and Mathematical Sciences, Nanyang Technological University, 21 Nanyang Link 637371, Singapore}
\email{christos@ntu.edu.sg}

\date{\today}
\begin{abstract}
We introduce a new class of primitive building blocks for realizing quantum logic elements based on nanoscale magnetization textures called skyrmions. In a skyrmion qubit, information is stored in the quantum degree of helicity, and the logical states can be adjusted by electric and magnetic fields, offering a rich operation regime with high anharmonicity. By exploring a large parameter space, we propose two skyrmion qubit variants depending on their quantized state. We discuss appropriate microwave pulses required to generate single-qubit gates for quantum computing, and skyrmion multiqubit schemes for a scalable architecture with tailored couplings. Scalability, controllability by microwave fields, operation time scales, and readout by nonvolatile techniques converge to make the skyrmion qubit highly attractive as a logical element of a quantum processor.
\end{abstract}

\maketitle

Quantum computing promises to dramatically improve computational power by harnessing the intrinsic properties of quantum mechanics. Its core is a quantum bit (qubit) of information made from a very small particle such as an atom, ion or electron. Proposed qubit systems include trapped atoms, quantum dots and photons\cite{Ladd2010,PhysRevA.57.120,PRXQuantum.2.010312}. Among them, superconducting circuits, currently one of the leading platforms for noisy intermediate-scale quantum computing protocols \cite{Preskill2018quantumcomputingin}, are macroscopic in size but with well-established quantum properties\cite{Clarke2008}. Nevertheless, despite tremendous progress, significant challenges remain, in particular with respect to  control and scalability \cite{PRXQuantum.2.017001}. 

Here we propose an alternative macroscopic qubit design based on magnetic skyrmions, topologically protected nanoscale magnetization textures, which have emerged as potential information carriers for future spintronic devices\cite{Bogdanov2020}. We focus on frustrated magnets, in which skyrmions and antiskyrmions have a new internal degree of freedom associated with the rotation of helicity \cite{PhysRevLett.108.017206,Leonov2015,PhysRevB.93.064430,Zhang2017,Leonov2017}. In these systems, the noncollinear spin texture induces electric polarization, allowing for electric-field modulation of the skyrmion helicity \cite{Matsukura2015,Yao_2020}. Along with magnetic field gradients \cite{Casiraghi2019} (MFGs) and microwave fields\cite{PhysRevLett.120.237203,Okamura2013}, electric fields emerge as a new, powerful tool for a current-free control of skyrmion dynamics \cite{Hsu2017}. Skyrmions of a few lattice sites\cite{Wiesendanger2016} inspired theoretical studies on their quantum properties \cite{PhysRevX.9.041063,PhysRevX.7.041045}. Similar to Josephson junctions\cite{PhysRevLett.55.1908,PhysRevLett.55.1543}, their macroscopic quantum tunneling and energy-level quantization are indicative of quantum behavior. In sufficiently small magnets, an analogous quantum behavior in terms of macroscopic quantum tunneling of the magnetic moment has been experimentally verified in mesoscopic magnetic systems \cite{PhysRevLett.68.3092,Thomas1996,Brooke2001}, while the quantum depinning of a magnetic skyrmion has been theoretically proposed \cite{PhysRevLett.124.097202}.
\begin{figure}[b]
  \includegraphics[width=\linewidth]{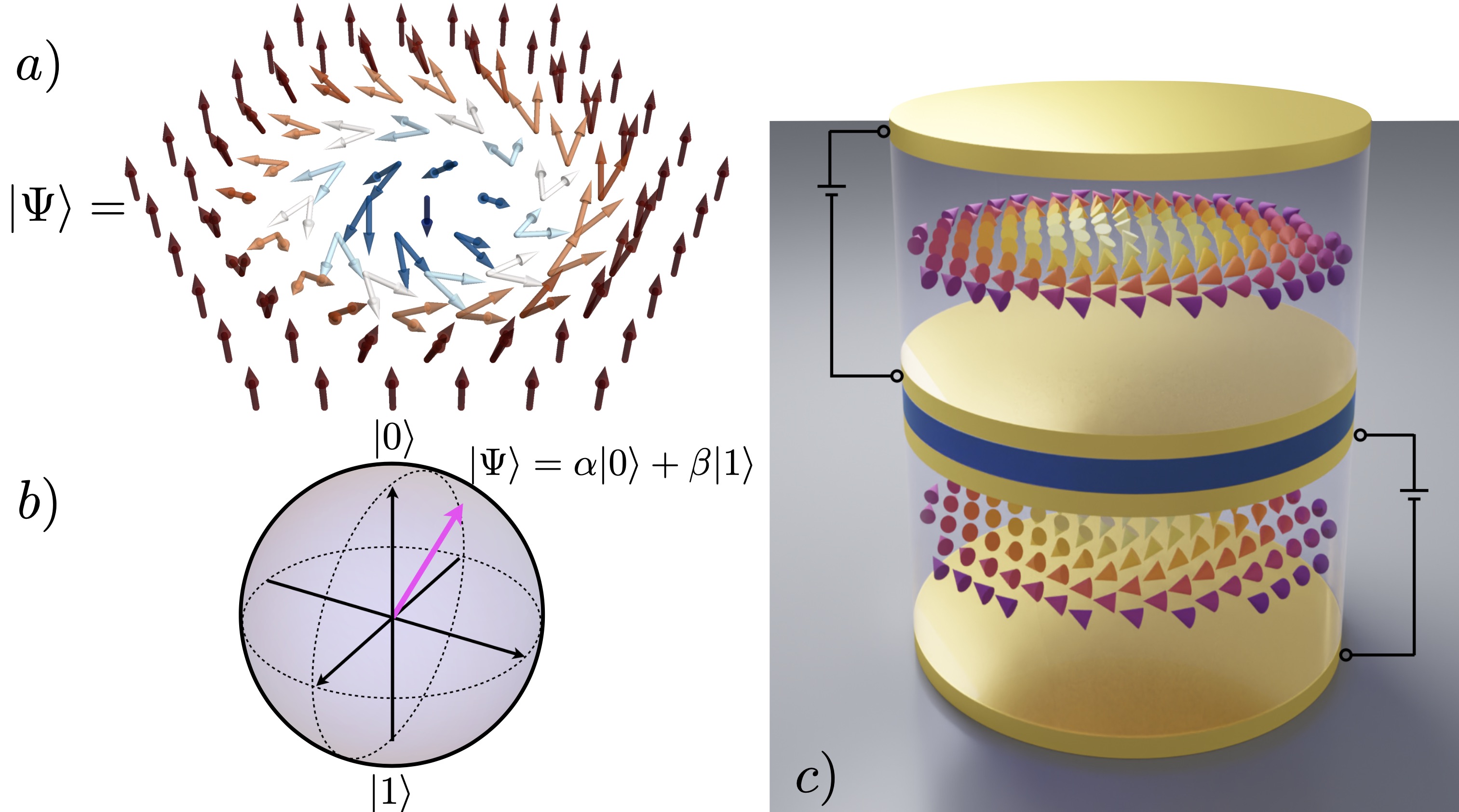} 
  \caption{Skyrmion Qubit Concept. (a) A quantum state $\vert \Psi \rangle$ as an arbitrary superposition of skyrmion configurations with distinct helicities $\vp$. (b) Bloch sphere representation of $\vert \Psi \rangle = \alpha \vert 0 \rangle + \beta \vert 1 \rangle$, with $\vert 0 \rangle$ and $\vert 1 \rangle$ denoting the two lowest energy levels of the quantum operator $\hvp$. (c) A bilayer of magnetic materials as a platform for the skyrmion qubit coupling scheme. The qubit coupling is tuned by a nonmagnetic spacer (blue), and logical states are adjusted by electric fields (yellow plates).}
  \label{Fig:skyrmionGates}
\end{figure}
\begin{figure*}[t]
  \includegraphics[width=\textwidth]{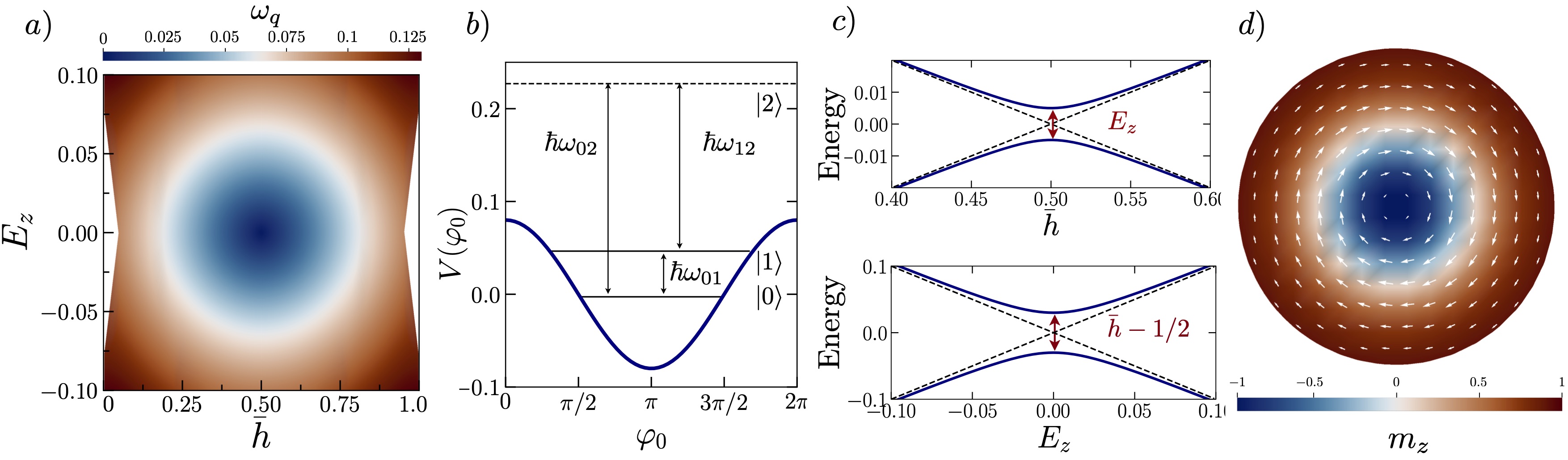}
  \caption{The $\boldsymbol{S_z}$-qubit properties. (a) Magnetic field $\bar{h}$ and electric field $E_z$ dependence of the transition frequency $\omega_q$, close to the degeneracy point $\bar{h}=0.5$. The colored surface represents the values of $\omega_q$ which satisfy the requirement of high anharmonicity. (b) Nonequidistant quantized energy levels and potential landscape. The qubit states are the ground state $\vert 0 \rangle$ and first excited state $\vert 1 \rangle$ with level spacing $\hbar \omega_{01}=\omega_q$ smaller than transitions to higher states $\hbar \omega_{02},\hbar \omega_{12}$.  (c) Universal energy level anticrossing diagram close to the degeneracy point (dashed lines). The degeneracy is lifted by an electric field (upper panel) or increasing the magnetic field away from $\bar{h}=0.5$ (lower panel). At the degeneracy point, energy eigenstates are symmetric and antisymmetric superpositions of the skyrmion qubit states $(\vert 0 \rangle \pm \vert 1 \rangle)/\sqrt{2}$.  (d) A magnetic skyrmion with a circular profile stabilized in a magnetic nanodisk.}
  \label{Fig:Sz}
\end{figure*} 
 ~\\
We formulate a theoretical framework of skyrmion quantization and construct skyrmion qubits based on the energy-level quantization of the helicity degree of freedom. The ability to control the energy-level spectra with external parameters, including electric and magnetic fields, offers a rich parameter space of possible qubit variants with high anharmonicity and tailored characteristics. We propose microwave MFGs for skyrmion qubit manipulation and gate operation, and consider skyrmion multiqubit schemes for a scalable architecture. A skyrmion qubit has a moderately-high coherence time in the microsecond regime, while nonvolatile readout techniques can be employed for a reliable qubit state readout. Finally, we discuss how scale-up multiqubit challenges can be addressed by leveraging state-of-the-art skyrmion technology and show that skyrmion qubits are suitable for quantum computing technology.

 \textit{\textbf{Skyrmion Field Quantization}}. We begin by considering the inversion-symmetric Heisenberg model with competing interactions \cite{PhysRevB.93.064430},
\begin{align}
\mc{F} = -\frac{J_1}{2} (\nabla \mb{m})^2 +\frac{J_2 a^2}{2} (\nabla^2 \mb{m})^2 -  \frac{H}{a^2} m_z + \frac{K}{a^2} m_z^2\,,
\label{eq:Energy}
\end{align}
where $H$ and $K$ are the Zeeman and anisotropy coupling, respectively, while $J_1$ and $J_2$ denote the strength of the competing interactions and $a$ the lattice spacing. A number of geometrically frustrated magnets are good candidates to host complex spin textures \cite{PhysRevLett.108.017206}, including the triangular-lattice magnet Gd$_2$PdSi$_3$, known to support skyrmion phases \cite{Kurumaji914}. Using $\mb{m}= [\sin \Theta \cos \Phi,\sin \Theta \sin \Phi, \cos \Theta]$, we describe classical skyrmions by $\Phi(\mb{r})=-Q \phi$ and $\Theta=\Theta(\rho)$, with $\rho,\phi$ polar coordinates. This class of solutions is characterized by an integer-valued topological charge $Q= (1/4 \pi) \int_{\mb{r}} 	\mb{m} \cdot (\pt_x \mb{m} \times \pt_y \mb{m})$, with $Q=1$ ($Q=-1$) for a skyrmion (antiskyrmion). The skyrmion size is defined as $\lambda \equiv 2 a	/\mbox{Re}[\gamma_\pm]$, with $\gamma_{\pm} = \sqrt{-1\pm \tilde{\gamma}}/\sqrt{2}$ and $\tilde{\gamma}=\sqrt{1-4(H/J_1+2K/J_1)}$. The model of Eq.~\eqref{eq:Energy} has an unbroken global symmetry, $\Phi \rightarrow \Phi + \vp$, with $\vp$ the collective coordinate of the skyrmion helicity. By considering a skyrmion stabilized in a nanodisk (see Fig.\ref{Fig:skyrmionGates}), we exclude the translational coordinate of position \cite{PhysRevX.7.041045} and focus exclusively on the dynamics of $\vp$.

To investigate quantum effects, we utilize a method of collective coordinate quantization. Here $\vp$ and its conjugate momentum $S_z$ are introduced by performing a canonical transformation in the phase space path integral \cite{PhysRevD.11.2943,PhysRevD.49.3598} (see Supplemental Material). This is achieved by ensuring momentum is conserved, $S_z=P$, with $P=\int_{\mb{r}}(1-\cos\Theta)\pt_{\phi} \Phi$ the infinitesimal generator of rotations satisfying $\{ P, \Phi \} = -\pt_{\phi} \Phi$. Using standard equivalence between path integral and canonical quantization, we introduce operators $\hvp$ and  $\hat{S}_z$ with $[\hvp, \hat{S}_z] = i/\bar{S}$, and $\bar{S}$ the effective spin. The classical limit is associated with $\bar{S} \gg 1$. Eigenstates of $\hat{S}_z$ are labeled by an integer charge $s$ with $\hat{S}_z \vert s \rangle = s/\bar{S} \vert s \rangle$, and states $\hvp \vert \vp \rangle = \vp \vert \vp \rangle$ have a circular topology $\vert \vp \rangle= \vert \vp +2 \pi \rangle$. The relation between physical and dimensionless parameters is summarized in Table.\ref{Units}. We construct skyrmion qubits based on textures with $Q=1$. Antiskyrmion qubits follow directly from our present analysis.  
\begin{figure*}[t]
  \includegraphics[width=\textwidth]{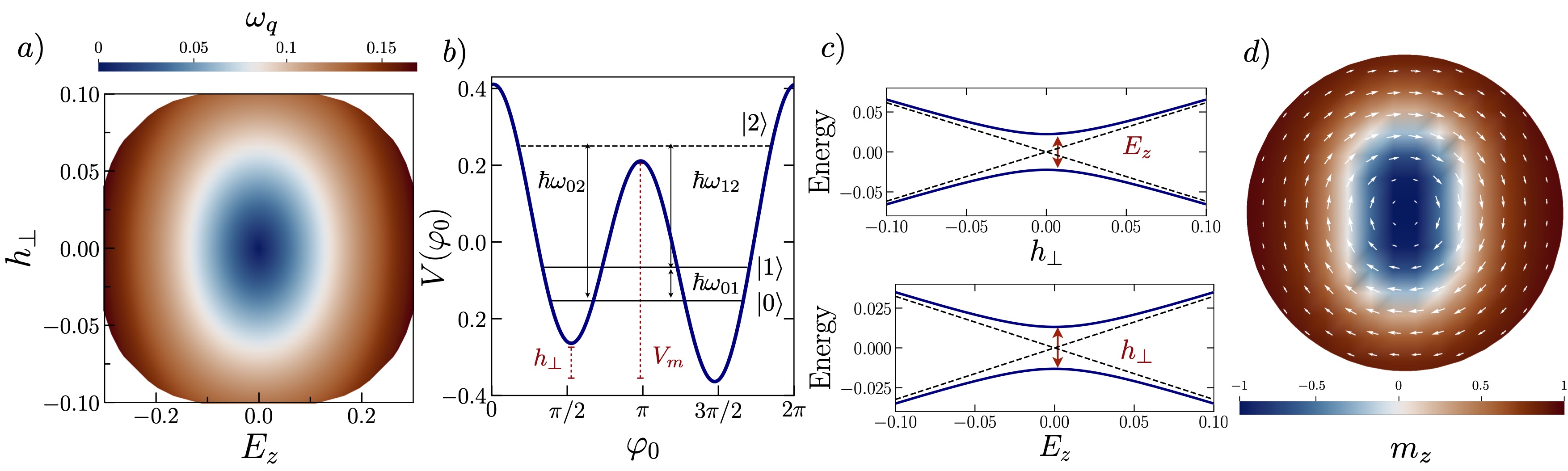}
  \caption{\textbf{The helicity-qubit properties.} a) Electric field $E_z$ and magnetic field gradient $h_\perp$ dependence of the transition frequency $\omega_q$, close to the degeneracy point $\bar{h}=1$. The colored surface represents the values of $\omega_q$ which satisfy the requirement of high anharmonicity. b) Nonequidistant quantized energy levels and double-well potential landscape. The qubit states are the ground state $\vert 0 \rangle$ and first excited state $\vert 1 \rangle$ with level spacing $\hbar \omega_{01}=\omega_q$ smaller than transitions to higher states $\hbar \omega_{02},\hbar \omega_{12}$. The potential barrier $V_m$ is controlled by $E_z$ and the well difference by $h_\perp$. c) Universal energy level anticrossing diagram close to the degeneracy point $\bar{h}=1$. The degeneracy is lifted by an electric field (upper panel) or a magnetic field gradient (lower panel). d) A magnetic skyrmion with an elliptical profile stabilized in a magnetic nanodisk. The elliptical profile is essential for realizing the double-well potential. }\label{Fig:Helicity}
\end{figure*}
~\\

 \textit{\textbf{Fundamental Skyrmion Qubit Types}}.
 We now seek to construct a skyrmion qubit based on the energy-level quantization of the helicity degree of freedom. A promising qubit candidate needs to satisfy several criteria including scalability, ability to initialize to a simple fiducial state, long decoherence times, a universal set of quantum gates, and the ability to perform qubit-specific measurements\cite{DiVincenzo2020}.
 
\noindent The $\boldsymbol{S_z}$--Qubit: The ability to control the energy-level spectra with external parameters, offers a rich parameter space of possible qubit variants with tailored characteristics. We introduce the $S_z$-qubit Hamiltonian,
\begin{align}
H_{S_z}= \kappa (\hat{S}_z - h/\kappa)^2 -E_z \cos \hvp\,,
\label{eq:SzHamiltonian}
\end{align}
which resembles the circuit Hamiltonian of a superconducting charge qubit \cite{doi:10.1146/annurev-conmatphys-031119-050605}. Here $\kappa$ and $h$ denote the anisotropy and magnetic field coupling, respectively, in dimensionless units. The noncollinear spin texture gives rise to an electric polarization which couples to an electric field $E_z$ applied across the nanodisk to control $\vp$\cite{Yao_2020} (see Fig.~\ref{Fig:skyrmionGates} for a schematic illustration of the setup). The $S_z$-qubit is designed in the $E_z \ll \kappa$ regime, such that logical qubits are spin states $\vert s \rangle$, representing deviations of the $m_z$ component from equilibrium. The solution of the Schr\"{o}dinger equation $H_{S_z} \Psi_s (\vp) =\mc{E}_s \Psi_s (\vp)$, with $\Psi_s (\vp) = \langle \vp \vert s \rangle$, can be calculated exactly in the form of special functions (see Supplemental Material). In Fig.~\ref{Fig:Sz}-(b) we plot the potential landscape and the first three levels using $\kappa=0.1$, $h=0.47$, and $E_z=0.02$.

Two requirements are essential for a reliable qubit operation; nonequidistance of the energy spectrum to uniquely address each transition and suppressed spontaneous thermal excitations to higher energy levels $k_B T \ll \hbar \omega_{12},\hbar \omega_{02}$. The remarkable feature of skyrmion qubits is that these conditions can be met by tuning the relevant external parameters. In Fig.~\ref{Fig:Sz}-(a) we present the range of parameters $\bar{h}=h \bar{S}/\kappa$ and $E_z$ for which a relatively large anharmonicity is present, $\vert \omega_{12}-\omega_{01} \vert > 20 \% \omega_{01} $ and $\vert \omega_{02}-\omega_{01} \vert > 20 \% \omega_{01}$. 

For $\bar{h}= 1/2$, the two lowest spin states $ \vert 0 \rangle$ and $\vert 1 \rangle$ are degenerate, and a small $E_z$ lifts the degeneracy creating a tight two-level system. Truncating the full Hilbert space to qubit subspace, the reduced Hamiltonian is
\begin{align}
H_q = \frac{H_0}{2} \hat{\sigma}_z - \frac{X_c}{2} \hat{\sigma}_x \,,
\label{eq:QubitHC}
\end{align}
with $H_0= \kappa (1-2 \bar{h})/\bar{S}$, $X_c=E_z$, and $\omega_q= \sqrt{H_0^2+X_c^2}$ the corresponding qubit level spacing. The universal level repulsion diagram is shown in Fig.~\ref{Fig:Sz}-c), with a minimum energy splitting $E_z$. The $S_z$-qubit operation regime in physical units is given in Table.~\ref{Lifetime}. We note that the proposed qubit platform has large anharmonicity, and the voltage bias for qubit manipulation is several orders of magnitude smaller compared to those required for the electric-field skyrmion creation and annihilation \cite{Hsu2017}. 

%%%%%%%%%%%%%%%%%%%%%%%%%%%%%%%%%%%%%%%%%%%%%%%%%%%
\begin{table*} \centering \caption{\label{Units} Relation between physical and dimensionless parameters. We use $J_1 =1$ meV, $a=5$ \AA, $\bar{S}=10$, $J_2=J_1$, $K=0.4 J_1$, $K_x=0.05 J_1$, and $P_E=20~\mu$C/cm$^2$. MFG stands for magnetic field gradient.}
\begin{ruledtabular}
\begin{tabular}{c|c|c|c|c|c|c}
 Length & Time  & Frequency  &Temperature &Magnetic field &Electric field & Static MFG \\
$\mb{r} \times 0.5$ nm &$t \times 6.6 ~~10^{-13}$ s& $\omega \times 1519$ GHz& $T \times 11.6$ K&$H/g\mu_B = h \times 0.86$ T& $E= E_z \times 215$ V/m & $H_\perp/g \mu_B = h_\perp \times 1.72$ T/nm\\  
\end{tabular}
\end{ruledtabular}
\end{table*}
%%%%%%%%%%%%%%%%%%%%%%%%%%%%%%%%%%%%%%%%%%%%%%%%%%%
\noindent The Helicity--Qubit: Inspired by the superconducting flux qubit and proposals on magnetic domain walls \cite{PhysRevB.97.064401}, we seek to construct a double-well potential landscape for the helicity $\vp$, in order to define the qubit logical space using the two well minima. This is achieved by considering a material with in-plane magnetic anisotropy of strength $\kappa_x$\cite{PhysRevB.99.094405} and a skyrmion characterized by an elliptical profile, as the result of defect engineering \cite{Arjana2020,Fernandes_2020}. The Hamiltonian for this new type of helicity qubit reads $H_{\vp}= \kappa \hat{S}_z -h \hat{S}_z +V(\hvp)$, with the double-well potential given by
\begin{align}
V(\vp) =\kappa_x \cos 2\hvp -E_z \cos \hvp + h_\perp \sin \hvp \,. 
\label{eq:Potential}
\end{align} 

The first two terms in Eq.~\eqref{eq:Potential} create a symmetric potential, and the third term describes a depth difference between the well created by an in-plane MFG of strength $h_{\perp}$. The solutions of the eigenvalue problem $H_{\vp} \Psi_n(\vp) = \mc{E}_n  \Psi_n(\vp)$ are $2\pi$-periodic functions calculated numerically. The potential in the helicity representation is schematically shown in Fig.~\ref{Fig:Helicity}-(b) together with the first three levels. Close to the degeneracy point at $\bar{h}=1$ and for $h_{\perp}=0$, the two lowest energy functions $\Psi_{0,1}$ are symmetric and antisymmetric combinations of the two wave functions localized in each well located at $\varphi_m=\tan^{-1}(\sqrt{16 \kappa_x^2-E_z^2}/E_z)$. A finite $h_\perp$ acts as an energy bias creating a depth well difference, such that the ground and first-excited states are now localized in different wells. 
%%%%%%%%%%%%%%%%%%%%%%%%%%%%%%%%%%%%%%%%%%%%%%%%%%%
\begin{table*} \centering \caption{\label{Lifetime} Skyrmion qubit operation regime and lifetime.  We use $\alpha=10^{-5}$ and $T=100$ mK. EF stands for electric field and MFG for magnetic field gradient. }
\begin{ruledtabular}
\begin{tabular}{cccccccc}
 Qubit type & Magnetic field& External control& $\omega_q$ &$T_1$& $T_2$& $\omega_{12}$ & $T_c$  \\ \hline \\ 
 $S_z$-qubit& $8.9$ mT&EF = $108$ mV/$\mu$m& 25.6 GHz& $0.27$ $\mu$s& $0.49$ $\mu$s& 310 GHz & 2.50 K \\ \hline  \\
 Helicity qubit& $445$ mT&EF = $296$ mV/$\mu$m& 14.9 GHz& $0.15$ $\mu$s& $0.26$ $\mu$s& 330 GHz & 2.60 K  \\ \hline  \\
 Helicity qubit& $445$ mT&MFG = 1.73 mT/nm& 2.1 GHz& $0.43$ $\mu$s& $0.32$ $\mu$s & 330 GHz & 2.55 K 
\end{tabular}
\end{ruledtabular}
\end{table*}

~\\

At $\bar{h}=1$, level anticrossing can be probed by applying either an electric field $E_z$ (see Fig.~\ref{Fig:Helicity}-(c) upper panel) or a magnetic field gradient $h_\perp$ (see Fig.~\ref{Fig:Helicity}-(c) lower panel). The reduced qubit Hamiltonian under the two-level approximation has the form of Eq.~\eqref{eq:QubitHC}, where $H_0 = \mc{E}_1-\mc{E}_0$ and $X_c = g_e E_z$ for $h_\perp=0$, or  $X_c = g_b h_\perp$ for $Ez=0$. Constants $H_0$, $g_e$, and $g_b$ are found numerically. The helicity-qubit operation regime in physical units is given in Table.~\ref{Lifetime}, using both $E_z$ and $h_\perp$ as external control parameters. 

\textit{\textbf{Qubit Control}}. A quantum coherent computation depends on the ability to control individual quantum degrees of freedom. Here we propose microwave MFGs for skyrmion qubit manipulation and gate operation. MFGs give rise to additional Hamiltonian terms $\Hext(t) = b f(t)\cos (\omega t+\pext) \cos \hvp$, with $f(t)$ a dimensionless envelope function, or in terms of the qubit Hamiltonian, $\Hext^q = b_x(t) \hat{\sigma}_x$, with $b_x(t)= b_0 f(t) \cos (\omega t+\pext)$. In the diagonal basis, the driven Hamiltonian is written as
\begin{align}
H_q = \frac{\omega_q}{2} \hat{\sigma}_z + b_x(t)[ \cos \theta \hat{\sigma}_x +\sin \theta \hat{\sigma}_z] \,, 
\end{align}
with $\tan \theta = X_c/H_0$. To elucidate the role of the drive, we transform $H_q$ into the rotating frame,
\begin{align}
\Hrot = \frac{\Delta \omega}{2} \hat{\sigma}_z +\frac{\Omega}{2}f(t)[\cos \pext \hat{\sigma}_x + \sin \pext \hat{\sigma}_y] \,,
\label{eq:HRot}
\end{align} 
where $\Delta \omega = \omega_q - \omega$ is the detuning frequency and $\Omega = b_0 \cos\theta$. Single-qubit operations correspond to rotations of the qubit state by a certain angle about a particular axis. As an example, for $\pext=0$ and $\Delta \omega=0$, the unitary operator $U_x(t) = e^{-\frac{i}{2}\vartheta(t) \hat{\sigma}_x}$ corresponds to rotations around the $x$ axis by an angle $\vartheta(t)= -\Omega \int_0^{t} f(t') dt'$ \cite{doi:10.1063/1.5089550}. Rotations about the $y$ axis are achieved for $\pext =\pi/2$. 

\textit{\textbf{Qubit Coupling Scheme}}. A key component for realizing a scalable quantum computer is an interaction Hamiltonian between individual qubits.  
As a straightforward scheme for coupling skyrmion qubits, we consider the interlayer exchange interaction in a magnetic bilayer mediated by a nonmagnetic spacer layer (see Fig.~\ref{Fig:skyrmionGates} for a visualization). The interaction term is given by $\Fint=\Jint \int_{\mb{r}} \mb{m}_1 \cdot \mb{m}_2$\cite{PhysRevB.81.104411}, or in terms of the helicities, $\Hint= -\Jint\cos(\varphi_1-\varphi_2)$. The resulting Hamiltonian in the qubit basis contains both transverse and longitudinal couplings, 
\begin{align}
\Hint = - \mJint^x \hat{\sigma}^1_x\hat{\sigma}^2_x - \mJint^z \hat{\sigma}^1_z\hat{\sigma}^2_z \,. 
\end{align}
$\Jint$ can be tuned experimentally by changing the spacer thickness, while both $\mJint^{x,z}$ allow for an independent control by tuning all three external fields $h$, $E_z$, and $h_\perp$. This property is especially important in applications where both longitudinal and transverse couplings are desired, such as quantum annealing \cite{doi:10.1063/1.5089550}. 

\textit{\textbf{Noise and Decoherence}}. The interaction of the skyrmion qubit with the environmental degrees of freedom is a source of noise that leads to decoherence. They result in Ohmic damping terms for the collective coordinates $\vp$ and $S_z$\cite{PhysRevLett.100.127204}, accompanied by random fluctuating forces $\xi_i$ that enter the quantum Hamiltonian as $\hat{H} \rightarrow \hat{H} + \xi_{\vp} \hvp + \xi_{S_z} \hat{S}_z$. $\xi_i$ is fully characterized by the classical ensemble averages $\langle \xi_i(t) \rangle=0$ and $\langle \xi_i(t) \xi_j(t') \rangle= \delta_{ij}S_i(t-t')$ \cite{PhysRevB.97.064401}, and the correlator $S_i(t)$ is defined via the fluctuation-dissipation theorem, $S_i(\omega) = \alpha_{i} \omega \coth(\beta \omega/2)$, with $\alpha_i$ constants proportional to the Gilbert damping $\alpha$. In terms of the reduced qubit Hamiltonian one finds,
\begin{align}
H_q &= \frac{\omega_q}{2} \hat{\sigma}_z + \xi_{x}(t) \gamma_{x} \hat{\sigma}_x +\xi_{y}(t) \gamma_{y} \hat{\sigma}_y+ \xi_{z}(t) \gamma_{z} \hat{\sigma}_z \,,
\end{align}
where $\gamma_{i}$ constants which depend on the qubit type and $\xi_{x,y,z}$ are linear combinations of $\xi_{\vp}$ and $\xi_{S_z}$. 

Within the Bloch-Redfield picture of two-level system dynamics, relaxation processes are characterized by the longitudinal relaxation rate $\Gamma_1=T_1^{-1}$ and the dephasing rate $\Gamma_2=T_2^{-1}$. The latter is a combination of effects of the depolarization $\Gamma_1$ and of the pure dephasing $\Gamma_\varphi$, combined to a rate $\Gamma_2= \Gamma_1/2+\Gamma_\varphi$, with $\Gamma_1 = \gamma_x^2 S_x(\omega_q) + \gamma_y^2 S_y(\omega_q)$ and $\Gamma_\varphi = \gamma_z^2 S_z(0)$\cite{PhysRevB.72.134519}. The optimal regime for realizing both long coherence and high anharmonicity is close to the degeneracy point and for $X_c \ll H_0$. This translates to the requirement $\bar{h}=0.5$ and $E_z \ll 1$ for the $S_z$-qubit, and to $\bar{h}=1$ and $E_z,h_\perp \ll 1$ for the helicity qubit. 

In Table.\ref{Lifetime} we present the expected qubit lifetimes for a modest choice of an ultralow Gilbert damping $\alpha=10^{-5}$ and $T = 100$ mK. A skyrmion qubit has a moderately high coherence time in the microsecond regime. This is comparable to early measurements of the flux superconducting qubit and 2 orders of magnitude larger than the Cooper pair box\cite{doi:10.1146/annurev-conmatphys-031119-050605}. The number of coherent Rabi frequency oscillations within the coherence time is $\Omega T_1 \propto 10^5$, inside the desired margins expected for superconducting qubits \cite{Devoret1169,PhysRevB.97.064401}. Several magnetic thin films exhibit ultralow Gilbert damping of the order of $\alpha \sim 10^{-4}-10^{-5}$~~\cite{Soumah2018,PhysRevApplied.11.064009,PhysRevLett.107.066604}. In the sub-Kelvin qubit operational regime, Gilbert damping is expected to be even lower \cite{PhysRevB.95.214423,Okada3815}. Coherence times can be further improved with the development of cleaner magnetic samples and interfaces in engineered architectures, without trading off qubit anharmonicity and scalability.

\textit{\textbf{Readout Techniques}}. An essential part for implementing skyrmion-based quantum-computing architectures is a reliable readout. Quantum sensing of coherent single-magnon techniques, based on quantum dot\cite{Jackson2021} or superconducting qubit\cite{Lachance-Quirion425} sensors, is promising for the readout of $S_z$-qubit states, single magnetic excitations from the equilibrium configuration. On the other hand, helicity-qubit states represent two distinct skyrmion configurations with helicity values located at the two minima of the double-well potential of Eq.\eqref{eq:Potential}. Experimental observation of skyrmion helicity is possible using nitrogen-vacancy (NV) magnetometry\cite{Dovzhenko2018}, allowing for a detector-single qubit coupling control by varying the NV sensor distance from the skyrmion. Resonant elastic x-ray scattering\cite{PhysRevLett.120.227202} techniques provide a direct observation of skyrmion helicity, and when combined with ferromagnetic resonance measurements \cite{PhysRevLett.123.167201} can offer a promising single-qubit readout method. Finally, coupling a skyrmion to a magnetic force microscopy resonator allows the detection of magnetic states, which appear as resonance frequency shift signals\cite{2103.10382}.

\textit{\textbf{Conclusions}}. We proposed a novel physical qubit platform based on magnetic nanoskyrmions in frustrated magnets. The skyrmion state, energy-level spectra, transition frequency and qubit lifetime are configurable and can be engineered by adjusting external electric and magnetic fields, offering a rich operation regime with high anharmonicity. Microwave pulses were shown to generate single-qubit gates for quantum computing, and skyrmion multiqubit schemes were considered for a scalable architecture with tailored couplings. Whereas, nonvolatile readout techniques can be employed for a reliable qubit state readout, using state-of-the-art magnetic sensing technology. We anticipate the considerable progress in the field of skyrmionics will provide exciting new directions on the development of skyrmion qubits as promising candidates for quantum computing technology.

\begin{acknowledgments} We thank Martino Poggio, So Takei, Daniel Loss, Ivar Martin and Markus Garst for useful discussions. C. Psaroudaki has received funding from the European Union’s Horizon 2020 research and innovation program under the Marie Sklodowska-Curie Grant Agreement No. 839004. C. Panagopoulos acknowledges support from the Singapore National Research Foundation (NRF) NRF-Investigatorship (No. NRFNRFI2015-04) and Singapore MOE Academic Research Fund Tier 3 Grant No.MOE2018-T3-1-002.
\end{acknowledgments} 

\bibliography{Skyrmion_Qubit}

%merlin.mbs apsrev4-1.bst 2010-07-25 4.21a (PWD, AO, DPC) hacked
%Control: key (0)
%Control: author (8) initials jnrlst
%Control: editor formatted (1) identically to author
%Control: production of article title (-1) disabled
%Control: page (0) single
%Control: year (1) truncated
%Control: production of eprint (0) enabled
\begin{thebibliography}{52}%
\makeatletter
\providecommand \@ifxundefined [1]{%
 \@ifx{#1\undefined}
}%
\providecommand \@ifnum [1]{%
 \ifnum #1\expandafter \@firstoftwo
 \else \expandafter \@secondoftwo
 \fi
}%
\providecommand \@ifx [1]{%
 \ifx #1\expandafter \@firstoftwo
 \else \expandafter \@secondoftwo
 \fi
}%
\providecommand \natexlab [1]{#1}%
\providecommand \enquote  [1]{``#1''}%
\providecommand \bibnamefont  [1]{#1}%
\providecommand \bibfnamefont [1]{#1}%
\providecommand \citenamefont [1]{#1}%
\providecommand \href@noop [0]{\@secondoftwo}%
\providecommand \href [0]{\begingroup \@sanitize@url \@href}%
\providecommand \@href[1]{\@@startlink{#1}\@@href}%
\providecommand \@@href[1]{\endgroup#1\@@endlink}%
\providecommand \@sanitize@url [0]{\catcode `\\12\catcode `\$12\catcode
  `\&12\catcode `\#12\catcode `\^12\catcode `\_12\catcode `\%12\relax}%
\providecommand \@@startlink[1]{}%
\providecommand \@@endlink[0]{}%
\providecommand \url  [0]{\begingroup\@sanitize@url \@url }%
\providecommand \@url [1]{\endgroup\@href {#1}{\urlprefix }}%
\providecommand \urlprefix  [0]{URL }%
\providecommand \Eprint [0]{\href }%
\providecommand \doibase [0]{http://dx.doi.org/}%
\providecommand \selectlanguage [0]{\@gobble}%
\providecommand \bibinfo  [0]{\@secondoftwo}%
\providecommand \bibfield  [0]{\@secondoftwo}%
\providecommand \translation [1]{[#1]}%
\providecommand \BibitemOpen [0]{}%
\providecommand \bibitemStop [0]{}%
\providecommand \bibitemNoStop [0]{.\EOS\space}%
\providecommand \EOS [0]{\spacefactor3000\relax}%
\providecommand \BibitemShut  [1]{\csname bibitem#1\endcsname}%
\let\auto@bib@innerbib\@empty
%</preamble>
\bibitem [{\citenamefont {Ladd}\ \emph {et~al.}(2010)\citenamefont {Ladd},
  \citenamefont {Jelezko}, \citenamefont {Laflamme}, \citenamefont {Nakamura},
  \citenamefont {Monroe},\ and\ \citenamefont {O'Brien}}]{Ladd2010}%
  \BibitemOpen
  \bibfield  {author} {\bibinfo {author} {\bibfnamefont {T.~D.}\ \bibnamefont
  {Ladd}}, \bibinfo {author} {\bibfnamefont {F.}~\bibnamefont {Jelezko}},
  \bibinfo {author} {\bibfnamefont {R.}~\bibnamefont {Laflamme}}, \bibinfo
  {author} {\bibfnamefont {Y.}~\bibnamefont {Nakamura}}, \bibinfo {author}
  {\bibfnamefont {C.}~\bibnamefont {Monroe}}, \ and\ \bibinfo {author}
  {\bibfnamefont {J.~L.}\ \bibnamefont {O'Brien}},\ }\href {\doibase
  10.1038/nature08812} {\bibfield  {journal} {\bibinfo  {journal} {Nature}\
  }\textbf {\bibinfo {volume} {464}},\ \bibinfo {pages} {45} (\bibinfo {year}
  {2010})}\BibitemShut {NoStop}%
\bibitem [{\citenamefont {Loss}\ and\ \citenamefont
  {DiVincenzo}(1998)}]{PhysRevA.57.120}%
  \BibitemOpen
  \bibfield  {author} {\bibinfo {author} {\bibfnamefont {D.}~\bibnamefont
  {Loss}}\ and\ \bibinfo {author} {\bibfnamefont {D.~P.}\ \bibnamefont
  {DiVincenzo}},\ }\href {\doibase 10.1103/PhysRevA.57.120} {\bibfield
  {journal} {\bibinfo  {journal} {Phys. Rev. A}\ }\textbf {\bibinfo {volume}
  {57}},\ \bibinfo {pages} {120} (\bibinfo {year} {1998})}\BibitemShut
  {NoStop}%
\bibitem [{\citenamefont {Grimm}\ \emph {et~al.}(2021)\citenamefont {Grimm},
  \citenamefont {Beckert}, \citenamefont {Aeppli},\ and\ \citenamefont
  {M\"uller}}]{PRXQuantum.2.010312}%
  \BibitemOpen
  \bibfield  {author} {\bibinfo {author} {\bibfnamefont {M.}~\bibnamefont
  {Grimm}}, \bibinfo {author} {\bibfnamefont {A.}~\bibnamefont {Beckert}},
  \bibinfo {author} {\bibfnamefont {G.}~\bibnamefont {Aeppli}}, \ and\ \bibinfo
  {author} {\bibfnamefont {M.}~\bibnamefont {M\"uller}},\ }\href {\doibase
  10.1103/PRXQuantum.2.010312} {\bibfield  {journal} {\bibinfo  {journal} {PRX
  Quantum}\ }\textbf {\bibinfo {volume} {2}},\ \bibinfo {pages} {010312}
  (\bibinfo {year} {2021})}\BibitemShut {NoStop}%
\bibitem [{\citenamefont {Preskill}(2018)}]{Preskill2018quantumcomputingin}%
  \BibitemOpen
  \bibfield  {author} {\bibinfo {author} {\bibfnamefont {J.}~\bibnamefont
  {Preskill}},\ }\href {\doibase 10.22331/q-2018-08-06-79} {\bibfield
  {journal} {\bibinfo  {journal} {{Quantum}}\ }\textbf {\bibinfo {volume}
  {2}},\ \bibinfo {pages} {79} (\bibinfo {year} {2018})}\BibitemShut {NoStop}%
\bibitem [{\citenamefont {Clarke}\ and\ \citenamefont
  {Wilhelm}(2008)}]{Clarke2008}%
  \BibitemOpen
  \bibfield  {author} {\bibinfo {author} {\bibfnamefont {J.}~\bibnamefont
  {Clarke}}\ and\ \bibinfo {author} {\bibfnamefont {F.~K.}\ \bibnamefont
  {Wilhelm}},\ }\href {\doibase 10.1038/nature07128} {\bibfield  {journal}
  {\bibinfo  {journal} {Nature}\ }\textbf {\bibinfo {volume} {453}},\ \bibinfo
  {pages} {1031} (\bibinfo {year} {2008})}\BibitemShut {NoStop}%
\bibitem [{\citenamefont {Alexeev}\ \emph {et~al.}(2021)\citenamefont
  {Alexeev}, \citenamefont {Bacon}, \citenamefont {Brown}, \citenamefont
  {Calderbank}, \citenamefont {Carr}, \citenamefont {Chong}, \citenamefont
  {DeMarco}, \citenamefont {Englund}, \citenamefont {Farhi}, \citenamefont
  {Fefferman}, \citenamefont {Gorshkov}, \citenamefont {Houck}, \citenamefont
  {Kim}, \citenamefont {Kimmel}, \citenamefont {Lange}, \citenamefont {Lloyd},
  \citenamefont {Lukin}, \citenamefont {Maslov}, \citenamefont {Maunz},
  \citenamefont {Monroe}, \citenamefont {Preskill}, \citenamefont {Roetteler},
  \citenamefont {Savage},\ and\ \citenamefont
  {Thompson}}]{PRXQuantum.2.017001}%
  \BibitemOpen
  \bibfield  {author} {\bibinfo {author} {\bibfnamefont {Y.}~\bibnamefont
  {Alexeev}}, \bibinfo {author} {\bibfnamefont {D.}~\bibnamefont {Bacon}},
  \bibinfo {author} {\bibfnamefont {K.~R.}\ \bibnamefont {Brown}}, \bibinfo
  {author} {\bibfnamefont {R.}~\bibnamefont {Calderbank}}, \bibinfo {author}
  {\bibfnamefont {L.~D.}\ \bibnamefont {Carr}}, \bibinfo {author}
  {\bibfnamefont {F.~T.}\ \bibnamefont {Chong}}, \bibinfo {author}
  {\bibfnamefont {B.}~\bibnamefont {DeMarco}}, \bibinfo {author} {\bibfnamefont
  {D.}~\bibnamefont {Englund}}, \bibinfo {author} {\bibfnamefont
  {E.}~\bibnamefont {Farhi}}, \bibinfo {author} {\bibfnamefont
  {B.}~\bibnamefont {Fefferman}}, \bibinfo {author} {\bibfnamefont {A.~V.}\
  \bibnamefont {Gorshkov}}, \bibinfo {author} {\bibfnamefont {A.}~\bibnamefont
  {Houck}}, \bibinfo {author} {\bibfnamefont {J.}~\bibnamefont {Kim}}, \bibinfo
  {author} {\bibfnamefont {S.}~\bibnamefont {Kimmel}}, \bibinfo {author}
  {\bibfnamefont {M.}~\bibnamefont {Lange}}, \bibinfo {author} {\bibfnamefont
  {S.}~\bibnamefont {Lloyd}}, \bibinfo {author} {\bibfnamefont {M.~D.}\
  \bibnamefont {Lukin}}, \bibinfo {author} {\bibfnamefont {D.}~\bibnamefont
  {Maslov}}, \bibinfo {author} {\bibfnamefont {P.}~\bibnamefont {Maunz}},
  \bibinfo {author} {\bibfnamefont {C.}~\bibnamefont {Monroe}}, \bibinfo
  {author} {\bibfnamefont {J.}~\bibnamefont {Preskill}}, \bibinfo {author}
  {\bibfnamefont {M.}~\bibnamefont {Roetteler}}, \bibinfo {author}
  {\bibfnamefont {M.~J.}\ \bibnamefont {Savage}}, \ and\ \bibinfo {author}
  {\bibfnamefont {J.}~\bibnamefont {Thompson}},\ }\href {\doibase
  10.1103/PRXQuantum.2.017001} {\bibfield  {journal} {\bibinfo  {journal} {PRX
  Quantum}\ }\textbf {\bibinfo {volume} {2}},\ \bibinfo {pages} {017001}
  (\bibinfo {year} {2021})}\BibitemShut {NoStop}%
\bibitem [{\citenamefont {Bogdanov}\ and\ \citenamefont
  {Panagopoulos}(2020)}]{Bogdanov2020}%
  \BibitemOpen
  \bibfield  {author} {\bibinfo {author} {\bibfnamefont {A.~N.}\ \bibnamefont
  {Bogdanov}}\ and\ \bibinfo {author} {\bibfnamefont {C.}~\bibnamefont
  {Panagopoulos}},\ }\href {\doibase 10.1038/s42254-020-0203-7} {\bibfield
  {journal} {\bibinfo  {journal} {Nature Reviews Physics}\ }\textbf {\bibinfo
  {volume} {2}},\ \bibinfo {pages} {492} (\bibinfo {year} {2020})}\BibitemShut
  {NoStop}%
\bibitem [{\citenamefont {Okubo}\ \emph {et~al.}(2012)\citenamefont {Okubo},
  \citenamefont {Chung},\ and\ \citenamefont
  {Kawamura}}]{PhysRevLett.108.017206}%
  \BibitemOpen
  \bibfield  {author} {\bibinfo {author} {\bibfnamefont {T.}~\bibnamefont
  {Okubo}}, \bibinfo {author} {\bibfnamefont {S.}~\bibnamefont {Chung}}, \ and\
  \bibinfo {author} {\bibfnamefont {H.}~\bibnamefont {Kawamura}},\ }\href
  {\doibase 10.1103/PhysRevLett.108.017206} {\bibfield  {journal} {\bibinfo
  {journal} {Phys. Rev. Lett.}\ }\textbf {\bibinfo {volume} {108}},\ \bibinfo
  {pages} {017206} (\bibinfo {year} {2012})}\BibitemShut {NoStop}%
\bibitem [{\citenamefont {Leonov}\ and\ \citenamefont
  {Mostovoy}(2015)}]{Leonov2015}%
  \BibitemOpen
  \bibfield  {author} {\bibinfo {author} {\bibfnamefont {A.~O.}\ \bibnamefont
  {Leonov}}\ and\ \bibinfo {author} {\bibfnamefont {M.}~\bibnamefont
  {Mostovoy}},\ }\href {\doibase 10.1038/ncomms9275} {\bibfield  {journal}
  {\bibinfo  {journal} {Nature Communications}\ }\textbf {\bibinfo {volume}
  {6}},\ \bibinfo {pages} {8275} (\bibinfo {year} {2015})}\BibitemShut
  {NoStop}%
\bibitem [{\citenamefont {Lin}\ and\ \citenamefont
  {Hayami}(2016)}]{PhysRevB.93.064430}%
  \BibitemOpen
  \bibfield  {author} {\bibinfo {author} {\bibfnamefont {S.-Z.}\ \bibnamefont
  {Lin}}\ and\ \bibinfo {author} {\bibfnamefont {S.}~\bibnamefont {Hayami}},\
  }\href {\doibase 10.1103/PhysRevB.93.064430} {\bibfield  {journal} {\bibinfo
  {journal} {Phys. Rev. B}\ }\textbf {\bibinfo {volume} {93}},\ \bibinfo
  {pages} {064430} (\bibinfo {year} {2016})}\BibitemShut {NoStop}%
\bibitem [{\citenamefont {Zhang}\ \emph {et~al.}(2017)\citenamefont {Zhang},
  \citenamefont {Xia}, \citenamefont {Zhou}, \citenamefont {Liu}, \citenamefont
  {Zhang},\ and\ \citenamefont {Ezawa}}]{Zhang2017}%
  \BibitemOpen
  \bibfield  {author} {\bibinfo {author} {\bibfnamefont {X.}~\bibnamefont
  {Zhang}}, \bibinfo {author} {\bibfnamefont {J.}~\bibnamefont {Xia}}, \bibinfo
  {author} {\bibfnamefont {Y.}~\bibnamefont {Zhou}}, \bibinfo {author}
  {\bibfnamefont {X.}~\bibnamefont {Liu}}, \bibinfo {author} {\bibfnamefont
  {H.}~\bibnamefont {Zhang}}, \ and\ \bibinfo {author} {\bibfnamefont
  {M.}~\bibnamefont {Ezawa}},\ }\href {\doibase 10.1038/s41467-017-01785-w}
  {\bibfield  {journal} {\bibinfo  {journal} {Nature Communications}\ }\textbf
  {\bibinfo {volume} {8}},\ \bibinfo {pages} {1717} (\bibinfo {year}
  {2017})}\BibitemShut {NoStop}%
\bibitem [{\citenamefont {Leonov}\ and\ \citenamefont
  {Mostovoy}(2017)}]{Leonov2017}%
  \BibitemOpen
  \bibfield  {author} {\bibinfo {author} {\bibfnamefont {A.~O.}\ \bibnamefont
  {Leonov}}\ and\ \bibinfo {author} {\bibfnamefont {M.}~\bibnamefont
  {Mostovoy}},\ }\href {\doibase 10.1038/ncomms14394} {\bibfield  {journal}
  {\bibinfo  {journal} {Nature Communications}\ }\textbf {\bibinfo {volume}
  {8}},\ \bibinfo {pages} {14394} (\bibinfo {year} {2017})}\BibitemShut
  {NoStop}%
\bibitem [{\citenamefont {Matsukura}\ \emph {et~al.}(2015)\citenamefont
  {Matsukura}, \citenamefont {Tokura},\ and\ \citenamefont
  {Ohno}}]{Matsukura2015}%
  \BibitemOpen
  \bibfield  {author} {\bibinfo {author} {\bibfnamefont {F.}~\bibnamefont
  {Matsukura}}, \bibinfo {author} {\bibfnamefont {Y.}~\bibnamefont {Tokura}}, \
  and\ \bibinfo {author} {\bibfnamefont {H.}~\bibnamefont {Ohno}},\ }\href
  {\doibase 10.1038/nnano.2015.22} {\bibfield  {journal} {\bibinfo  {journal}
  {Nature Nanotechnology}\ }\textbf {\bibinfo {volume} {10}},\ \bibinfo {pages}
  {209} (\bibinfo {year} {2015})}\BibitemShut {NoStop}%
\bibitem [{\citenamefont {Yao}\ \emph {et~al.}(2020)\citenamefont {Yao},
  \citenamefont {Chen},\ and\ \citenamefont {Dong}}]{Yao_2020}%
  \BibitemOpen
  \bibfield  {author} {\bibinfo {author} {\bibfnamefont {X.}~\bibnamefont
  {Yao}}, \bibinfo {author} {\bibfnamefont {J.}~\bibnamefont {Chen}}, \ and\
  \bibinfo {author} {\bibfnamefont {S.}~\bibnamefont {Dong}},\ }\href {\doibase
  10.1088/1367-2630/aba1b3} {\bibfield  {journal} {\bibinfo  {journal} {New
  Journal of Physics}\ }\textbf {\bibinfo {volume} {22}},\ \bibinfo {pages}
  {083032} (\bibinfo {year} {2020})}\BibitemShut {NoStop}%
\bibitem [{\citenamefont {Casiraghi}\ \emph {et~al.}(2019)\citenamefont
  {Casiraghi}, \citenamefont {Corte-Le{\'o}n}, \citenamefont {Vafaee},
  \citenamefont {Garcia-Sanchez}, \citenamefont {Durin}, \citenamefont
  {Pasquale}, \citenamefont {Jakob}, \citenamefont {Kl{\"a}ui},\ and\
  \citenamefont {Kazakova}}]{Casiraghi2019}%
  \BibitemOpen
  \bibfield  {author} {\bibinfo {author} {\bibfnamefont {A.}~\bibnamefont
  {Casiraghi}}, \bibinfo {author} {\bibfnamefont {H.}~\bibnamefont
  {Corte-Le{\'o}n}}, \bibinfo {author} {\bibfnamefont {M.}~\bibnamefont
  {Vafaee}}, \bibinfo {author} {\bibfnamefont {F.}~\bibnamefont
  {Garcia-Sanchez}}, \bibinfo {author} {\bibfnamefont {G.}~\bibnamefont
  {Durin}}, \bibinfo {author} {\bibfnamefont {M.}~\bibnamefont {Pasquale}},
  \bibinfo {author} {\bibfnamefont {G.}~\bibnamefont {Jakob}}, \bibinfo
  {author} {\bibfnamefont {M.}~\bibnamefont {Kl{\"a}ui}}, \ and\ \bibinfo
  {author} {\bibfnamefont {O.}~\bibnamefont {Kazakova}},\ }\href {\doibase
  10.1038/s42005-019-0242-5} {\bibfield  {journal} {\bibinfo  {journal}
  {Communications Physics}\ }\textbf {\bibinfo {volume} {2}},\ \bibinfo {pages}
  {145} (\bibinfo {year} {2019})}\BibitemShut {NoStop}%
\bibitem [{\citenamefont {Psaroudaki}\ and\ \citenamefont
  {Loss}(2018)}]{PhysRevLett.120.237203}%
  \BibitemOpen
  \bibfield  {author} {\bibinfo {author} {\bibfnamefont {C.}~\bibnamefont
  {Psaroudaki}}\ and\ \bibinfo {author} {\bibfnamefont {D.}~\bibnamefont
  {Loss}},\ }\href {\doibase 10.1103/PhysRevLett.120.237203} {\bibfield
  {journal} {\bibinfo  {journal} {Phys. Rev. Lett.}\ }\textbf {\bibinfo
  {volume} {120}},\ \bibinfo {pages} {237203} (\bibinfo {year}
  {2018})}\BibitemShut {NoStop}%
\bibitem [{\citenamefont {Okamura}\ \emph {et~al.}(2013)\citenamefont
  {Okamura}, \citenamefont {Kagawa}, \citenamefont {Mochizuki}, \citenamefont
  {Kubota}, \citenamefont {Seki}, \citenamefont {Ishiwata}, \citenamefont
  {Kawasaki}, \citenamefont {Onose},\ and\ \citenamefont
  {Tokura}}]{Okamura2013}%
  \BibitemOpen
  \bibfield  {author} {\bibinfo {author} {\bibfnamefont {Y.}~\bibnamefont
  {Okamura}}, \bibinfo {author} {\bibfnamefont {F.}~\bibnamefont {Kagawa}},
  \bibinfo {author} {\bibfnamefont {M.}~\bibnamefont {Mochizuki}}, \bibinfo
  {author} {\bibfnamefont {M.}~\bibnamefont {Kubota}}, \bibinfo {author}
  {\bibfnamefont {S.}~\bibnamefont {Seki}}, \bibinfo {author} {\bibfnamefont
  {S.}~\bibnamefont {Ishiwata}}, \bibinfo {author} {\bibfnamefont
  {M.}~\bibnamefont {Kawasaki}}, \bibinfo {author} {\bibfnamefont
  {Y.}~\bibnamefont {Onose}}, \ and\ \bibinfo {author} {\bibfnamefont
  {Y.}~\bibnamefont {Tokura}},\ }\href {\doibase 10.1038/ncomms3391} {\bibfield
   {journal} {\bibinfo  {journal} {Nature Communications}\ }\textbf {\bibinfo
  {volume} {4}},\ \bibinfo {pages} {2391} (\bibinfo {year} {2013})}\BibitemShut
  {NoStop}%
\bibitem [{\citenamefont {Hsu}\ \emph {et~al.}(2017)\citenamefont {Hsu},
  \citenamefont {Kubetzka}, \citenamefont {Finco}, \citenamefont {Romming},
  \citenamefont {von Bergmann},\ and\ \citenamefont {Wiesendanger}}]{Hsu2017}%
  \BibitemOpen
  \bibfield  {author} {\bibinfo {author} {\bibfnamefont {P.-J.}\ \bibnamefont
  {Hsu}}, \bibinfo {author} {\bibfnamefont {A.}~\bibnamefont {Kubetzka}},
  \bibinfo {author} {\bibfnamefont {A.}~\bibnamefont {Finco}}, \bibinfo
  {author} {\bibfnamefont {N.}~\bibnamefont {Romming}}, \bibinfo {author}
  {\bibfnamefont {K.}~\bibnamefont {von Bergmann}}, \ and\ \bibinfo {author}
  {\bibfnamefont {R.}~\bibnamefont {Wiesendanger}},\ }\href {\doibase
  10.1038/nnano.2016.234} {\bibfield  {journal} {\bibinfo  {journal} {Nature
  Nanotechnology}\ }\textbf {\bibinfo {volume} {12}},\ \bibinfo {pages} {123}
  (\bibinfo {year} {2017})}\BibitemShut {NoStop}%
\bibitem [{\citenamefont {Wiesendanger}(2016)}]{Wiesendanger2016}%
  \BibitemOpen
  \bibfield  {author} {\bibinfo {author} {\bibfnamefont {R.}~\bibnamefont
  {Wiesendanger}},\ }\href {\doibase 10.1038/natrevmats.2016.44} {\bibfield
  {journal} {\bibinfo  {journal} {Nature Reviews Materials}\ }\textbf {\bibinfo
  {volume} {1}},\ \bibinfo {pages} {16044} (\bibinfo {year}
  {2016})}\BibitemShut {NoStop}%
\bibitem [{\citenamefont {Lohani}\ \emph {et~al.}(2019)\citenamefont {Lohani},
  \citenamefont {Hickey}, \citenamefont {Masell},\ and\ \citenamefont
  {Rosch}}]{PhysRevX.9.041063}%
  \BibitemOpen
  \bibfield  {author} {\bibinfo {author} {\bibfnamefont {V.}~\bibnamefont
  {Lohani}}, \bibinfo {author} {\bibfnamefont {C.}~\bibnamefont {Hickey}},
  \bibinfo {author} {\bibfnamefont {J.}~\bibnamefont {Masell}}, \ and\ \bibinfo
  {author} {\bibfnamefont {A.}~\bibnamefont {Rosch}},\ }\href {\doibase
  10.1103/PhysRevX.9.041063} {\bibfield  {journal} {\bibinfo  {journal} {Phys.
  Rev. X}\ }\textbf {\bibinfo {volume} {9}},\ \bibinfo {pages} {041063}
  (\bibinfo {year} {2019})}\BibitemShut {NoStop}%
\bibitem [{\citenamefont {Psaroudaki}\ \emph {et~al.}(2017)\citenamefont
  {Psaroudaki}, \citenamefont {Hoffman}, \citenamefont {Klinovaja},\ and\
  \citenamefont {Loss}}]{PhysRevX.7.041045}%
  \BibitemOpen
  \bibfield  {author} {\bibinfo {author} {\bibfnamefont {C.}~\bibnamefont
  {Psaroudaki}}, \bibinfo {author} {\bibfnamefont {S.}~\bibnamefont {Hoffman}},
  \bibinfo {author} {\bibfnamefont {J.}~\bibnamefont {Klinovaja}}, \ and\
  \bibinfo {author} {\bibfnamefont {D.}~\bibnamefont {Loss}},\ }\href {\doibase
  10.1103/PhysRevX.7.041045} {\bibfield  {journal} {\bibinfo  {journal} {Phys.
  Rev. X}\ }\textbf {\bibinfo {volume} {7}},\ \bibinfo {pages} {041045}
  (\bibinfo {year} {2017})}\BibitemShut {NoStop}%
\bibitem [{\citenamefont {Devoret}\ \emph {et~al.}(1985)\citenamefont
  {Devoret}, \citenamefont {Martinis},\ and\ \citenamefont
  {Clarke}}]{PhysRevLett.55.1908}%
  \BibitemOpen
  \bibfield  {author} {\bibinfo {author} {\bibfnamefont {M.~H.}\ \bibnamefont
  {Devoret}}, \bibinfo {author} {\bibfnamefont {J.~M.}\ \bibnamefont
  {Martinis}}, \ and\ \bibinfo {author} {\bibfnamefont {J.}~\bibnamefont
  {Clarke}},\ }\href {\doibase 10.1103/PhysRevLett.55.1908} {\bibfield
  {journal} {\bibinfo  {journal} {Phys. Rev. Lett.}\ }\textbf {\bibinfo
  {volume} {55}},\ \bibinfo {pages} {1908} (\bibinfo {year}
  {1985})}\BibitemShut {NoStop}%
\bibitem [{\citenamefont {Martinis}\ \emph {et~al.}(1985)\citenamefont
  {Martinis}, \citenamefont {Devoret},\ and\ \citenamefont
  {Clarke}}]{PhysRevLett.55.1543}%
  \BibitemOpen
  \bibfield  {author} {\bibinfo {author} {\bibfnamefont {J.~M.}\ \bibnamefont
  {Martinis}}, \bibinfo {author} {\bibfnamefont {M.~H.}\ \bibnamefont
  {Devoret}}, \ and\ \bibinfo {author} {\bibfnamefont {J.}~\bibnamefont
  {Clarke}},\ }\href {\doibase 10.1103/PhysRevLett.55.1543} {\bibfield
  {journal} {\bibinfo  {journal} {Phys. Rev. Lett.}\ }\textbf {\bibinfo
  {volume} {55}},\ \bibinfo {pages} {1543} (\bibinfo {year}
  {1985})}\BibitemShut {NoStop}%
\bibitem [{\citenamefont {Awschalom}\ \emph {et~al.}(1992)\citenamefont
  {Awschalom}, \citenamefont {Smyth}, \citenamefont {Grinstein}, \citenamefont
  {DiVincenzo},\ and\ \citenamefont {Loss}}]{PhysRevLett.68.3092}%
  \BibitemOpen
  \bibfield  {author} {\bibinfo {author} {\bibfnamefont {D.~D.}\ \bibnamefont
  {Awschalom}}, \bibinfo {author} {\bibfnamefont {J.~F.}\ \bibnamefont
  {Smyth}}, \bibinfo {author} {\bibfnamefont {G.}~\bibnamefont {Grinstein}},
  \bibinfo {author} {\bibfnamefont {D.~P.}\ \bibnamefont {DiVincenzo}}, \ and\
  \bibinfo {author} {\bibfnamefont {D.}~\bibnamefont {Loss}},\ }\href {\doibase
  10.1103/PhysRevLett.68.3092} {\bibfield  {journal} {\bibinfo  {journal}
  {Phys. Rev. Lett.}\ }\textbf {\bibinfo {volume} {68}},\ \bibinfo {pages}
  {3092} (\bibinfo {year} {1992})}\BibitemShut {NoStop}%
\bibitem [{\citenamefont {Thomas}\ \emph {et~al.}(1996)\citenamefont {Thomas},
  \citenamefont {Lionti}, \citenamefont {Ballou}, \citenamefont {Gatteschi},
  \citenamefont {Sessoli},\ and\ \citenamefont {Barbara}}]{Thomas1996}%
  \BibitemOpen
  \bibfield  {author} {\bibinfo {author} {\bibfnamefont {L.}~\bibnamefont
  {Thomas}}, \bibinfo {author} {\bibfnamefont {F.}~\bibnamefont {Lionti}},
  \bibinfo {author} {\bibfnamefont {R.}~\bibnamefont {Ballou}}, \bibinfo
  {author} {\bibfnamefont {D.}~\bibnamefont {Gatteschi}}, \bibinfo {author}
  {\bibfnamefont {R.}~\bibnamefont {Sessoli}}, \ and\ \bibinfo {author}
  {\bibfnamefont {B.}~\bibnamefont {Barbara}},\ }\href {\doibase
  10.1038/383145a0} {\bibfield  {journal} {\bibinfo  {journal} {Nature}\
  }\textbf {\bibinfo {volume} {383}},\ \bibinfo {pages} {145} (\bibinfo {year}
  {1996})}\BibitemShut {NoStop}%
\bibitem [{\citenamefont {Brooke}\ \emph {et~al.}(2001)\citenamefont {Brooke},
  \citenamefont {Rosenbaum},\ and\ \citenamefont {Aeppli}}]{Brooke2001}%
  \BibitemOpen
  \bibfield  {author} {\bibinfo {author} {\bibfnamefont {J.}~\bibnamefont
  {Brooke}}, \bibinfo {author} {\bibfnamefont {T.~F.}\ \bibnamefont
  {Rosenbaum}}, \ and\ \bibinfo {author} {\bibfnamefont {G.}~\bibnamefont
  {Aeppli}},\ }\href {\doibase 10.1038/35098037} {\bibfield  {journal}
  {\bibinfo  {journal} {Nature}\ }\textbf {\bibinfo {volume} {413}},\ \bibinfo
  {pages} {610} (\bibinfo {year} {2001})}\BibitemShut {NoStop}%
\bibitem [{\citenamefont {Psaroudaki}\ and\ \citenamefont
  {Loss}(2020)}]{PhysRevLett.124.097202}%
  \BibitemOpen
  \bibfield  {author} {\bibinfo {author} {\bibfnamefont {C.}~\bibnamefont
  {Psaroudaki}}\ and\ \bibinfo {author} {\bibfnamefont {D.}~\bibnamefont
  {Loss}},\ }\href {\doibase 10.1103/PhysRevLett.124.097202} {\bibfield
  {journal} {\bibinfo  {journal} {Phys. Rev. Lett.}\ }\textbf {\bibinfo
  {volume} {124}},\ \bibinfo {pages} {097202} (\bibinfo {year}
  {2020})}\BibitemShut {NoStop}%
\bibitem [{\citenamefont {Kurumaji}\ \emph {et~al.}(2019)\citenamefont
  {Kurumaji}, \citenamefont {Nakajima}, \citenamefont {Hirschberger},
  \citenamefont {Kikkawa}, \citenamefont {Yamasaki}, \citenamefont {Sagayama},
  \citenamefont {Nakao}, \citenamefont {Taguchi}, \citenamefont {Arima},\ and\
  \citenamefont {Tokura}}]{Kurumaji914}%
  \BibitemOpen
  \bibfield  {author} {\bibinfo {author} {\bibfnamefont {T.}~\bibnamefont
  {Kurumaji}}, \bibinfo {author} {\bibfnamefont {T.}~\bibnamefont {Nakajima}},
  \bibinfo {author} {\bibfnamefont {M.}~\bibnamefont {Hirschberger}}, \bibinfo
  {author} {\bibfnamefont {A.}~\bibnamefont {Kikkawa}}, \bibinfo {author}
  {\bibfnamefont {Y.}~\bibnamefont {Yamasaki}}, \bibinfo {author}
  {\bibfnamefont {H.}~\bibnamefont {Sagayama}}, \bibinfo {author}
  {\bibfnamefont {H.}~\bibnamefont {Nakao}}, \bibinfo {author} {\bibfnamefont
  {Y.}~\bibnamefont {Taguchi}}, \bibinfo {author} {\bibfnamefont {T.-h.}\
  \bibnamefont {Arima}}, \ and\ \bibinfo {author} {\bibfnamefont
  {Y.}~\bibnamefont {Tokura}},\ }\href {\doibase 10.1126/science.aau0968}
  {\bibfield  {journal} {\bibinfo  {journal} {Science}\ }\textbf {\bibinfo
  {volume} {365}},\ \bibinfo {pages} {914} (\bibinfo {year}
  {2019})}\BibitemShut {NoStop}%
\bibitem [{\citenamefont {Gervais}\ and\ \citenamefont
  {Sakita}(1975)}]{PhysRevD.11.2943}%
  \BibitemOpen
  \bibfield  {author} {\bibinfo {author} {\bibfnamefont {J.~L.}\ \bibnamefont
  {Gervais}}\ and\ \bibinfo {author} {\bibfnamefont {B.}~\bibnamefont
  {Sakita}},\ }\href {\doibase 10.1103/PhysRevD.11.2943} {\bibfield  {journal}
  {\bibinfo  {journal} {Phys. Rev. D}\ }\textbf {\bibinfo {volume} {11}},\
  \bibinfo {pages} {2943} (\bibinfo {year} {1975})}\BibitemShut {NoStop}%
\bibitem [{\citenamefont {Dorey}\ \emph {et~al.}(1994)\citenamefont {Dorey},
  \citenamefont {Hughes},\ and\ \citenamefont {Mattis}}]{PhysRevD.49.3598}%
  \BibitemOpen
  \bibfield  {author} {\bibinfo {author} {\bibfnamefont {N.}~\bibnamefont
  {Dorey}}, \bibinfo {author} {\bibfnamefont {J.}~\bibnamefont {Hughes}}, \
  and\ \bibinfo {author} {\bibfnamefont {M.~P.}\ \bibnamefont {Mattis}},\
  }\href {\doibase 10.1103/PhysRevD.49.3598} {\bibfield  {journal} {\bibinfo
  {journal} {Phys. Rev. D}\ }\textbf {\bibinfo {volume} {49}},\ \bibinfo
  {pages} {3598} (\bibinfo {year} {1994})}\BibitemShut {NoStop}%
\bibitem [{\citenamefont {DiVincenzo}(2000)}]{DiVincenzo2020}%
  \BibitemOpen
  \bibfield  {author} {\bibinfo {author} {\bibfnamefont {D.~P.}\ \bibnamefont
  {DiVincenzo}},\ }\href {\doibase
  https://doi.org/10.1002/1521-3978(200009)48:9/11<771::AID-PROP771>3.0.CO;2-E}
  {\bibfield  {journal} {\bibinfo  {journal} {Fortschritte der Physik}\
  }\textbf {\bibinfo {volume} {48}},\ \bibinfo {pages} {771} (\bibinfo {year}
  {2000})}\BibitemShut {NoStop}%
\bibitem [{\citenamefont {Kjaergaard}\ \emph {et~al.}(2020)\citenamefont
  {Kjaergaard}, \citenamefont {Schwartz}, \citenamefont {Braumüller},
  \citenamefont {Krantz}, \citenamefont {Wang}, \citenamefont {Gustavsson},\
  and\ \citenamefont {Oliver}}]{doi:10.1146/annurev-conmatphys-031119-050605}%
  \BibitemOpen
  \bibfield  {author} {\bibinfo {author} {\bibfnamefont {M.}~\bibnamefont
  {Kjaergaard}}, \bibinfo {author} {\bibfnamefont {M.~E.}\ \bibnamefont
  {Schwartz}}, \bibinfo {author} {\bibfnamefont {J.}~\bibnamefont
  {Braumüller}}, \bibinfo {author} {\bibfnamefont {P.}~\bibnamefont {Krantz}},
  \bibinfo {author} {\bibfnamefont {J.~I.-J.}\ \bibnamefont {Wang}}, \bibinfo
  {author} {\bibfnamefont {S.}~\bibnamefont {Gustavsson}}, \ and\ \bibinfo
  {author} {\bibfnamefont {W.~D.}\ \bibnamefont {Oliver}},\ }\href {\doibase
  10.1146/annurev-conmatphys-031119-050605} {\bibfield  {journal} {\bibinfo
  {journal} {Annual Review of Condensed Matter Physics}\ }\textbf {\bibinfo
  {volume} {11}},\ \bibinfo {pages} {369} (\bibinfo {year} {2020})}\BibitemShut
  {NoStop}%
\bibitem [{\citenamefont {Takei}\ and\ \citenamefont
  {Mohseni}(2018)}]{PhysRevB.97.064401}%
  \BibitemOpen
  \bibfield  {author} {\bibinfo {author} {\bibfnamefont {S.}~\bibnamefont
  {Takei}}\ and\ \bibinfo {author} {\bibfnamefont {M.}~\bibnamefont
  {Mohseni}},\ }\href {\doibase 10.1103/PhysRevB.97.064401} {\bibfield
  {journal} {\bibinfo  {journal} {Phys. Rev. B}\ }\textbf {\bibinfo {volume}
  {97}},\ \bibinfo {pages} {064401} (\bibinfo {year} {2018})}\BibitemShut
  {NoStop}%
\bibitem [{\citenamefont {Roy}\ \emph {et~al.}(2019)\citenamefont {Roy},
  \citenamefont {Otxoa},\ and\ \citenamefont {Moutafis}}]{PhysRevB.99.094405}%
  \BibitemOpen
  \bibfield  {author} {\bibinfo {author} {\bibfnamefont {P.~E.}\ \bibnamefont
  {Roy}}, \bibinfo {author} {\bibfnamefont {R.~M.}\ \bibnamefont {Otxoa}}, \
  and\ \bibinfo {author} {\bibfnamefont {C.}~\bibnamefont {Moutafis}},\ }\href
  {\doibase 10.1103/PhysRevB.99.094405} {\bibfield  {journal} {\bibinfo
  {journal} {Phys. Rev. B}\ }\textbf {\bibinfo {volume} {99}},\ \bibinfo
  {pages} {094405} (\bibinfo {year} {2019})}\BibitemShut {NoStop}%
\bibitem [{\citenamefont {Arjana}\ \emph {et~al.}(2020)\citenamefont {Arjana},
  \citenamefont {Lima~Fernandes}, \citenamefont {Chico},\ and\ \citenamefont
  {Lounis}}]{Arjana2020}%
  \BibitemOpen
  \bibfield  {author} {\bibinfo {author} {\bibfnamefont {I.~G.}\ \bibnamefont
  {Arjana}}, \bibinfo {author} {\bibfnamefont {I.}~\bibnamefont
  {Lima~Fernandes}}, \bibinfo {author} {\bibfnamefont {J.}~\bibnamefont
  {Chico}}, \ and\ \bibinfo {author} {\bibfnamefont {S.}~\bibnamefont
  {Lounis}},\ }\href {\doibase 10.1038/s41598-020-71232-2} {\bibfield
  {journal} {\bibinfo  {journal} {Scientific Reports}\ }\textbf {\bibinfo
  {volume} {10}},\ \bibinfo {pages} {14655} (\bibinfo {year}
  {2020})}\BibitemShut {NoStop}%
\bibitem [{\citenamefont {Fernandes}\ \emph {et~al.}(2020)\citenamefont
  {Fernandes}, \citenamefont {Chico},\ and\ \citenamefont
  {Lounis}}]{Fernandes_2020}%
  \BibitemOpen
  \bibfield  {author} {\bibinfo {author} {\bibfnamefont {I.~L.}\ \bibnamefont
  {Fernandes}}, \bibinfo {author} {\bibfnamefont {J.}~\bibnamefont {Chico}}, \
  and\ \bibinfo {author} {\bibfnamefont {S.}~\bibnamefont {Lounis}},\ }\href
  {\doibase 10.1088/1361-648x/ab9cf0} {\bibfield  {journal} {\bibinfo
  {journal} {Journal of Physics: Condensed Matter}\ }\textbf {\bibinfo {volume}
  {32}},\ \bibinfo {pages} {425802} (\bibinfo {year} {2020})}\BibitemShut
  {NoStop}%
\bibitem [{\citenamefont {Krantz}\ \emph {et~al.}(2019)\citenamefont {Krantz},
  \citenamefont {Kjaergaard}, \citenamefont {Yan}, \citenamefont {Orlando},
  \citenamefont {Gustavsson},\ and\ \citenamefont
  {Oliver}}]{doi:10.1063/1.5089550}%
  \BibitemOpen
  \bibfield  {author} {\bibinfo {author} {\bibfnamefont {P.}~\bibnamefont
  {Krantz}}, \bibinfo {author} {\bibfnamefont {M.}~\bibnamefont {Kjaergaard}},
  \bibinfo {author} {\bibfnamefont {F.}~\bibnamefont {Yan}}, \bibinfo {author}
  {\bibfnamefont {T.~P.}\ \bibnamefont {Orlando}}, \bibinfo {author}
  {\bibfnamefont {S.}~\bibnamefont {Gustavsson}}, \ and\ \bibinfo {author}
  {\bibfnamefont {W.~D.}\ \bibnamefont {Oliver}},\ }\href {\doibase
  10.1063/1.5089550} {\bibfield  {journal} {\bibinfo  {journal} {Applied
  Physics Reviews}\ }\textbf {\bibinfo {volume} {6}},\ \bibinfo {pages}
  {021318} (\bibinfo {year} {2019})}\BibitemShut {NoStop}%
\bibitem [{\citenamefont {Poienar}\ \emph {et~al.}(2010)\citenamefont
  {Poienar}, \citenamefont {Damay}, \citenamefont {Martin}, \citenamefont
  {Robert},\ and\ \citenamefont {Petit}}]{PhysRevB.81.104411}%
  \BibitemOpen
  \bibfield  {author} {\bibinfo {author} {\bibfnamefont {M.}~\bibnamefont
  {Poienar}}, \bibinfo {author} {\bibfnamefont {F.}~\bibnamefont {Damay}},
  \bibinfo {author} {\bibfnamefont {C.}~\bibnamefont {Martin}}, \bibinfo
  {author} {\bibfnamefont {J.}~\bibnamefont {Robert}}, \ and\ \bibinfo {author}
  {\bibfnamefont {S.}~\bibnamefont {Petit}},\ }\href {\doibase
  10.1103/PhysRevB.81.104411} {\bibfield  {journal} {\bibinfo  {journal} {Phys.
  Rev. B}\ }\textbf {\bibinfo {volume} {81}},\ \bibinfo {pages} {104411}
  (\bibinfo {year} {2010})}\BibitemShut {NoStop}%
\bibitem [{\citenamefont {Tretiakov}\ \emph {et~al.}(2008)\citenamefont
  {Tretiakov}, \citenamefont {Clarke}, \citenamefont {Chern}, \citenamefont
  {Bazaliy},\ and\ \citenamefont {Tchernyshyov}}]{PhysRevLett.100.127204}%
  \BibitemOpen
  \bibfield  {author} {\bibinfo {author} {\bibfnamefont {O.~A.}\ \bibnamefont
  {Tretiakov}}, \bibinfo {author} {\bibfnamefont {D.}~\bibnamefont {Clarke}},
  \bibinfo {author} {\bibfnamefont {G.-W.}\ \bibnamefont {Chern}}, \bibinfo
  {author} {\bibfnamefont {Y.~B.}\ \bibnamefont {Bazaliy}}, \ and\ \bibinfo
  {author} {\bibfnamefont {O.}~\bibnamefont {Tchernyshyov}},\ }\href {\doibase
  10.1103/PhysRevLett.100.127204} {\bibfield  {journal} {\bibinfo  {journal}
  {Phys. Rev. Lett.}\ }\textbf {\bibinfo {volume} {100}},\ \bibinfo {pages}
  {127204} (\bibinfo {year} {2008})}\BibitemShut {NoStop}%
\bibitem [{\citenamefont {Ithier}\ \emph {et~al.}(2005)\citenamefont {Ithier},
  \citenamefont {Collin}, \citenamefont {Joyez}, \citenamefont {Meeson},
  \citenamefont {Vion}, \citenamefont {Esteve}, \citenamefont {Chiarello},
  \citenamefont {Shnirman}, \citenamefont {Makhlin}, \citenamefont {Schriefl},\
  and\ \citenamefont {Sch\"on}}]{PhysRevB.72.134519}%
  \BibitemOpen
  \bibfield  {author} {\bibinfo {author} {\bibfnamefont {G.}~\bibnamefont
  {Ithier}}, \bibinfo {author} {\bibfnamefont {E.}~\bibnamefont {Collin}},
  \bibinfo {author} {\bibfnamefont {P.}~\bibnamefont {Joyez}}, \bibinfo
  {author} {\bibfnamefont {P.~J.}\ \bibnamefont {Meeson}}, \bibinfo {author}
  {\bibfnamefont {D.}~\bibnamefont {Vion}}, \bibinfo {author} {\bibfnamefont
  {D.}~\bibnamefont {Esteve}}, \bibinfo {author} {\bibfnamefont
  {F.}~\bibnamefont {Chiarello}}, \bibinfo {author} {\bibfnamefont
  {A.}~\bibnamefont {Shnirman}}, \bibinfo {author} {\bibfnamefont
  {Y.}~\bibnamefont {Makhlin}}, \bibinfo {author} {\bibfnamefont
  {J.}~\bibnamefont {Schriefl}}, \ and\ \bibinfo {author} {\bibfnamefont
  {G.}~\bibnamefont {Sch\"on}},\ }\href {\doibase 10.1103/PhysRevB.72.134519}
  {\bibfield  {journal} {\bibinfo  {journal} {Phys. Rev. B}\ }\textbf {\bibinfo
  {volume} {72}},\ \bibinfo {pages} {134519} (\bibinfo {year}
  {2005})}\BibitemShut {NoStop}%
\bibitem [{\citenamefont {Devoret}\ and\ \citenamefont
  {Schoelkopf}(2013)}]{Devoret1169}%
  \BibitemOpen
  \bibfield  {author} {\bibinfo {author} {\bibfnamefont {M.~H.}\ \bibnamefont
  {Devoret}}\ and\ \bibinfo {author} {\bibfnamefont {R.~J.}\ \bibnamefont
  {Schoelkopf}},\ }\href {\doibase 10.1126/science.1231930} {\bibfield
  {journal} {\bibinfo  {journal} {Science}\ }\textbf {\bibinfo {volume}
  {339}},\ \bibinfo {pages} {1169} (\bibinfo {year} {2013})}\BibitemShut
  {NoStop}%
\bibitem [{\citenamefont {Soumah}\ \emph {et~al.}(2018)\citenamefont {Soumah},
  \citenamefont {Beaulieu}, \citenamefont {Qassym}, \citenamefont
  {Carr{\'e}t{\'e}ro}, \citenamefont {Jacquet}, \citenamefont {Lebourgeois},
  \citenamefont {Ben~Youssef}, \citenamefont {Bortolotti}, \citenamefont
  {Cros},\ and\ \citenamefont {Anane}}]{Soumah2018}%
  \BibitemOpen
  \bibfield  {author} {\bibinfo {author} {\bibfnamefont {L.}~\bibnamefont
  {Soumah}}, \bibinfo {author} {\bibfnamefont {N.}~\bibnamefont {Beaulieu}},
  \bibinfo {author} {\bibfnamefont {L.}~\bibnamefont {Qassym}}, \bibinfo
  {author} {\bibfnamefont {C.}~\bibnamefont {Carr{\'e}t{\'e}ro}}, \bibinfo
  {author} {\bibfnamefont {E.}~\bibnamefont {Jacquet}}, \bibinfo {author}
  {\bibfnamefont {R.}~\bibnamefont {Lebourgeois}}, \bibinfo {author}
  {\bibfnamefont {J.}~\bibnamefont {Ben~Youssef}}, \bibinfo {author}
  {\bibfnamefont {P.}~\bibnamefont {Bortolotti}}, \bibinfo {author}
  {\bibfnamefont {V.}~\bibnamefont {Cros}}, \ and\ \bibinfo {author}
  {\bibfnamefont {A.}~\bibnamefont {Anane}},\ }\href {\doibase
  10.1038/s41467-018-05732-1} {\bibfield  {journal} {\bibinfo  {journal}
  {Nature Communications}\ }\textbf {\bibinfo {volume} {9}},\ \bibinfo {pages}
  {3355} (\bibinfo {year} {2018})}\BibitemShut {NoStop}%
\bibitem [{\citenamefont {Guillemard}\ \emph {et~al.}(2019)\citenamefont
  {Guillemard}, \citenamefont {Petit-Watelot}, \citenamefont {Pasquier},
  \citenamefont {Pierre}, \citenamefont {Ghanbaja}, \citenamefont
  {Rojas-S\'anchez}, \citenamefont {Bataille}, \citenamefont {Rault},
  \citenamefont {Le~F\`evre}, \citenamefont {Bertran},\ and\ \citenamefont
  {Andrieu}}]{PhysRevApplied.11.064009}%
  \BibitemOpen
  \bibfield  {author} {\bibinfo {author} {\bibfnamefont {C.}~\bibnamefont
  {Guillemard}}, \bibinfo {author} {\bibfnamefont {S.}~\bibnamefont
  {Petit-Watelot}}, \bibinfo {author} {\bibfnamefont {L.}~\bibnamefont
  {Pasquier}}, \bibinfo {author} {\bibfnamefont {D.}~\bibnamefont {Pierre}},
  \bibinfo {author} {\bibfnamefont {J.}~\bibnamefont {Ghanbaja}}, \bibinfo
  {author} {\bibfnamefont {J.-C.}\ \bibnamefont {Rojas-S\'anchez}}, \bibinfo
  {author} {\bibfnamefont {A.}~\bibnamefont {Bataille}}, \bibinfo {author}
  {\bibfnamefont {J.}~\bibnamefont {Rault}}, \bibinfo {author} {\bibfnamefont
  {P.}~\bibnamefont {Le~F\`evre}}, \bibinfo {author} {\bibfnamefont
  {F.}~\bibnamefont {Bertran}}, \ and\ \bibinfo {author} {\bibfnamefont
  {S.}~\bibnamefont {Andrieu}},\ }\href {\doibase
  10.1103/PhysRevApplied.11.064009} {\bibfield  {journal} {\bibinfo  {journal}
  {Phys. Rev. Applied}\ }\textbf {\bibinfo {volume} {11}},\ \bibinfo {pages}
  {064009} (\bibinfo {year} {2019})}\BibitemShut {NoStop}%
\bibitem [{\citenamefont {Heinrich}\ \emph {et~al.}(2011)\citenamefont
  {Heinrich}, \citenamefont {Burrowes}, \citenamefont {Montoya}, \citenamefont
  {Kardasz}, \citenamefont {Girt}, \citenamefont {Song}, \citenamefont {Sun},\
  and\ \citenamefont {Wu}}]{PhysRevLett.107.066604}%
  \BibitemOpen
  \bibfield  {author} {\bibinfo {author} {\bibfnamefont {B.}~\bibnamefont
  {Heinrich}}, \bibinfo {author} {\bibfnamefont {C.}~\bibnamefont {Burrowes}},
  \bibinfo {author} {\bibfnamefont {E.}~\bibnamefont {Montoya}}, \bibinfo
  {author} {\bibfnamefont {B.}~\bibnamefont {Kardasz}}, \bibinfo {author}
  {\bibfnamefont {E.}~\bibnamefont {Girt}}, \bibinfo {author} {\bibfnamefont
  {Y.-Y.}\ \bibnamefont {Song}}, \bibinfo {author} {\bibfnamefont
  {Y.}~\bibnamefont {Sun}}, \ and\ \bibinfo {author} {\bibfnamefont
  {M.}~\bibnamefont {Wu}},\ }\href {\doibase 10.1103/PhysRevLett.107.066604}
  {\bibfield  {journal} {\bibinfo  {journal} {Phys. Rev. Lett.}\ }\textbf
  {\bibinfo {volume} {107}},\ \bibinfo {pages} {066604} (\bibinfo {year}
  {2011})}\BibitemShut {NoStop}%
\bibitem [{\citenamefont {Maier-Flaig}\ \emph {et~al.}(2017)\citenamefont
  {Maier-Flaig}, \citenamefont {Klingler}, \citenamefont {Dubs}, \citenamefont
  {Surzhenko}, \citenamefont {Gross}, \citenamefont {Weiler}, \citenamefont
  {Huebl},\ and\ \citenamefont {Goennenwein}}]{PhysRevB.95.214423}%
  \BibitemOpen
  \bibfield  {author} {\bibinfo {author} {\bibfnamefont {H.}~\bibnamefont
  {Maier-Flaig}}, \bibinfo {author} {\bibfnamefont {S.}~\bibnamefont
  {Klingler}}, \bibinfo {author} {\bibfnamefont {C.}~\bibnamefont {Dubs}},
  \bibinfo {author} {\bibfnamefont {O.}~\bibnamefont {Surzhenko}}, \bibinfo
  {author} {\bibfnamefont {R.}~\bibnamefont {Gross}}, \bibinfo {author}
  {\bibfnamefont {M.}~\bibnamefont {Weiler}}, \bibinfo {author} {\bibfnamefont
  {H.}~\bibnamefont {Huebl}}, \ and\ \bibinfo {author} {\bibfnamefont
  {S.~T.~B.}\ \bibnamefont {Goennenwein}},\ }\href {\doibase
  10.1103/PhysRevB.95.214423} {\bibfield  {journal} {\bibinfo  {journal} {Phys.
  Rev. B}\ }\textbf {\bibinfo {volume} {95}},\ \bibinfo {pages} {214423}
  (\bibinfo {year} {2017})}\BibitemShut {NoStop}%
\bibitem [{\citenamefont {Okada}\ \emph {et~al.}(2017)\citenamefont {Okada},
  \citenamefont {He}, \citenamefont {Gu}, \citenamefont {Kanai}, \citenamefont
  {Soumyanarayanan}, \citenamefont {Lim}, \citenamefont {Tran}, \citenamefont
  {Mori}, \citenamefont {Maekawa}, \citenamefont {Matsukura}, \citenamefont
  {Ohno},\ and\ \citenamefont {Panagopoulos}}]{Okada3815}%
  \BibitemOpen
  \bibfield  {author} {\bibinfo {author} {\bibfnamefont {A.}~\bibnamefont
  {Okada}}, \bibinfo {author} {\bibfnamefont {S.}~\bibnamefont {He}}, \bibinfo
  {author} {\bibfnamefont {B.}~\bibnamefont {Gu}}, \bibinfo {author}
  {\bibfnamefont {S.}~\bibnamefont {Kanai}}, \bibinfo {author} {\bibfnamefont
  {A.}~\bibnamefont {Soumyanarayanan}}, \bibinfo {author} {\bibfnamefont
  {S.~T.}\ \bibnamefont {Lim}}, \bibinfo {author} {\bibfnamefont
  {M.}~\bibnamefont {Tran}}, \bibinfo {author} {\bibfnamefont {M.}~\bibnamefont
  {Mori}}, \bibinfo {author} {\bibfnamefont {S.}~\bibnamefont {Maekawa}},
  \bibinfo {author} {\bibfnamefont {F.}~\bibnamefont {Matsukura}}, \bibinfo
  {author} {\bibfnamefont {H.}~\bibnamefont {Ohno}}, \ and\ \bibinfo {author}
  {\bibfnamefont {C.}~\bibnamefont {Panagopoulos}},\ }\href {\doibase
  10.1073/pnas.1613864114} {\bibfield  {journal} {\bibinfo  {journal}
  {Proceedings of the National Academy of Sciences}\ }\textbf {\bibinfo
  {volume} {114}},\ \bibinfo {pages} {3815} (\bibinfo {year}
  {2017})}\BibitemShut {NoStop}%
\bibitem [{\citenamefont {Jackson}\ \emph {et~al.}(2021)\citenamefont
  {Jackson}, \citenamefont {Gangloff}, \citenamefont {Bodey}, \citenamefont
  {Zaporski}, \citenamefont {Bachorz}, \citenamefont {Clarke}, \citenamefont
  {Hugues}, \citenamefont {Le~Gall},\ and\ \citenamefont
  {Atat{\"u}re}}]{Jackson2021}%
  \BibitemOpen
  \bibfield  {author} {\bibinfo {author} {\bibfnamefont {D.~M.}\ \bibnamefont
  {Jackson}}, \bibinfo {author} {\bibfnamefont {D.~A.}\ \bibnamefont
  {Gangloff}}, \bibinfo {author} {\bibfnamefont {J.~H.}\ \bibnamefont {Bodey}},
  \bibinfo {author} {\bibfnamefont {L.}~\bibnamefont {Zaporski}}, \bibinfo
  {author} {\bibfnamefont {C.}~\bibnamefont {Bachorz}}, \bibinfo {author}
  {\bibfnamefont {E.}~\bibnamefont {Clarke}}, \bibinfo {author} {\bibfnamefont
  {M.}~\bibnamefont {Hugues}}, \bibinfo {author} {\bibfnamefont
  {C.}~\bibnamefont {Le~Gall}}, \ and\ \bibinfo {author} {\bibfnamefont
  {M.}~\bibnamefont {Atat{\"u}re}},\ }\href {\doibase
  10.1038/s41567-020-01161-4} {\bibfield  {journal} {\bibinfo  {journal}
  {Nature Physics}\ }\textbf {\bibinfo {volume} {17}},\ \bibinfo {pages} {585}
  (\bibinfo {year} {2021})}\BibitemShut {NoStop}%
\bibitem [{\citenamefont {Lachance-Quirion}\ \emph {et~al.}(2020)\citenamefont
  {Lachance-Quirion}, \citenamefont {Wolski}, \citenamefont {Tabuchi},
  \citenamefont {Kono}, \citenamefont {Usami},\ and\ \citenamefont
  {Nakamura}}]{Lachance-Quirion425}%
  \BibitemOpen
  \bibfield  {author} {\bibinfo {author} {\bibfnamefont {D.}~\bibnamefont
  {Lachance-Quirion}}, \bibinfo {author} {\bibfnamefont {S.~P.}\ \bibnamefont
  {Wolski}}, \bibinfo {author} {\bibfnamefont {Y.}~\bibnamefont {Tabuchi}},
  \bibinfo {author} {\bibfnamefont {S.}~\bibnamefont {Kono}}, \bibinfo {author}
  {\bibfnamefont {K.}~\bibnamefont {Usami}}, \ and\ \bibinfo {author}
  {\bibfnamefont {Y.}~\bibnamefont {Nakamura}},\ }\href {\doibase
  10.1126/science.aaz9236} {\bibfield  {journal} {\bibinfo  {journal}
  {Science}\ }\textbf {\bibinfo {volume} {367}},\ \bibinfo {pages} {425}
  (\bibinfo {year} {2020})}\BibitemShut {NoStop}%
\bibitem [{\citenamefont {Dovzhenko}\ \emph {et~al.}(2018)\citenamefont
  {Dovzhenko}, \citenamefont {Casola}, \citenamefont {Schlotter}, \citenamefont
  {Zhou}, \citenamefont {B{\"u}ttner}, \citenamefont {Walsworth}, \citenamefont
  {Beach},\ and\ \citenamefont {Yacoby}}]{Dovzhenko2018}%
  \BibitemOpen
  \bibfield  {author} {\bibinfo {author} {\bibfnamefont {Y.}~\bibnamefont
  {Dovzhenko}}, \bibinfo {author} {\bibfnamefont {F.}~\bibnamefont {Casola}},
  \bibinfo {author} {\bibfnamefont {S.}~\bibnamefont {Schlotter}}, \bibinfo
  {author} {\bibfnamefont {T.~X.}\ \bibnamefont {Zhou}}, \bibinfo {author}
  {\bibfnamefont {F.}~\bibnamefont {B{\"u}ttner}}, \bibinfo {author}
  {\bibfnamefont {R.~L.}\ \bibnamefont {Walsworth}}, \bibinfo {author}
  {\bibfnamefont {G.~S.~D.}\ \bibnamefont {Beach}}, \ and\ \bibinfo {author}
  {\bibfnamefont {A.}~\bibnamefont {Yacoby}},\ }\href {\doibase
  10.1038/s41467-018-05158-9} {\bibfield  {journal} {\bibinfo  {journal}
  {Nature Communications}\ }\textbf {\bibinfo {volume} {9}},\ \bibinfo {pages}
  {2712} (\bibinfo {year} {2018})}\BibitemShut {NoStop}%
\bibitem [{\citenamefont {Zhang}\ \emph {et~al.}(2018)\citenamefont {Zhang},
  \citenamefont {van~der Laan}, \citenamefont {Wang}, \citenamefont
  {Haghighirad},\ and\ \citenamefont {Hesjedal}}]{PhysRevLett.120.227202}%
  \BibitemOpen
  \bibfield  {author} {\bibinfo {author} {\bibfnamefont {S.~L.}\ \bibnamefont
  {Zhang}}, \bibinfo {author} {\bibfnamefont {G.}~\bibnamefont {van~der Laan}},
  \bibinfo {author} {\bibfnamefont {W.~W.}\ \bibnamefont {Wang}}, \bibinfo
  {author} {\bibfnamefont {A.~A.}\ \bibnamefont {Haghighirad}}, \ and\ \bibinfo
  {author} {\bibfnamefont {T.}~\bibnamefont {Hesjedal}},\ }\href {\doibase
  10.1103/PhysRevLett.120.227202} {\bibfield  {journal} {\bibinfo  {journal}
  {Phys. Rev. Lett.}\ }\textbf {\bibinfo {volume} {120}},\ \bibinfo {pages}
  {227202} (\bibinfo {year} {2018})}\BibitemShut {NoStop}%
\bibitem [{\citenamefont {P\"ollath}\ \emph {et~al.}(2019)\citenamefont
  {P\"ollath}, \citenamefont {Aqeel}, \citenamefont {Bauer}, \citenamefont
  {Luo}, \citenamefont {Ryll}, \citenamefont {Radu}, \citenamefont
  {Pfleiderer}, \citenamefont {Woltersdorf},\ and\ \citenamefont
  {Back}}]{PhysRevLett.123.167201}%
  \BibitemOpen
  \bibfield  {author} {\bibinfo {author} {\bibfnamefont {S.}~\bibnamefont
  {P\"ollath}}, \bibinfo {author} {\bibfnamefont {A.}~\bibnamefont {Aqeel}},
  \bibinfo {author} {\bibfnamefont {A.}~\bibnamefont {Bauer}}, \bibinfo
  {author} {\bibfnamefont {C.}~\bibnamefont {Luo}}, \bibinfo {author}
  {\bibfnamefont {H.}~\bibnamefont {Ryll}}, \bibinfo {author} {\bibfnamefont
  {F.}~\bibnamefont {Radu}}, \bibinfo {author} {\bibfnamefont {C.}~\bibnamefont
  {Pfleiderer}}, \bibinfo {author} {\bibfnamefont {G.}~\bibnamefont
  {Woltersdorf}}, \ and\ \bibinfo {author} {\bibfnamefont {C.~H.}\ \bibnamefont
  {Back}},\ }\href {\doibase 10.1103/PhysRevLett.123.167201} {\bibfield
  {journal} {\bibinfo  {journal} {Phys. Rev. Lett.}\ }\textbf {\bibinfo
  {volume} {123}},\ \bibinfo {pages} {167201} (\bibinfo {year}
  {2019})}\BibitemShut {NoStop}%
\bibitem [{\citenamefont {Marchiori}\ \emph {et~al.}(2021)\citenamefont
  {Marchiori}, \citenamefont {Ceccarelli}, \citenamefont {Rossi}, \citenamefont
  {Lorenzelli}, \citenamefont {Degen},\ and\ \citenamefont
  {Poggio}}]{2103.10382}%
  \BibitemOpen
  \bibfield  {author} {\bibinfo {author} {\bibfnamefont {E.}~\bibnamefont
  {Marchiori}}, \bibinfo {author} {\bibfnamefont {L.}~\bibnamefont
  {Ceccarelli}}, \bibinfo {author} {\bibfnamefont {N.}~\bibnamefont {Rossi}},
  \bibinfo {author} {\bibfnamefont {L.}~\bibnamefont {Lorenzelli}}, \bibinfo
  {author} {\bibfnamefont {C.~L.}\ \bibnamefont {Degen}}, \ and\ \bibinfo
  {author} {\bibfnamefont {M.}~\bibnamefont {Poggio}},\ }\href@noop {}
  {\enquote {\bibinfo {title} {Technical review: Imaging weak magnetic field
  patterns on the nanometer-scale and its application to 2d materials},}\ }
  (\bibinfo {year} {2021}),\ \Eprint {http://arxiv.org/abs/arXiv:2103.10382}
  {arXiv:2103.10382} \BibitemShut {NoStop}%
\end{thebibliography}%

\newpage
\section*{Supplemental Material for:\\ S\lowercase{kyrmion} Q\lowercase{ubits:}\\ A N\lowercase{ew} C\lowercase{lass of} Q\lowercase{uantum} L\lowercase{ogic} E\lowercase{lements} B\lowercase{ased on} N\lowercase{anoscale} M\lowercase{agnetization}}

\section{The model}

\sloppy We consider a thin magnetic insulator with normalized magnetization $\mb{m}= [\sin \Theta \cos \Phi,\sin \Theta \sin \Phi, \cos \Theta]$, described by the real time action
\begin{align}
\mc{S}=\bar{S} \int d\td{t} \int d\mb{\td{r}} [ \frac{S N_A}{\alpha}  \dot{\Phi}(\Pi-1) - N_A \mc{F} (\Phi,\Pi)] \,, 
\label{eq:Action}
\end{align}
where $\bar{S}$ is the magnitude of the effective spin and $a$ is the lattice spacing. $\dot{\Phi}= \partial_{\td{t}} \Phi$ denotes the real time derivative, and $\Pi = \cos \Theta$ is canonically conjugate to $\Phi$. We consider the inversion-symmetric classical Heisenberg model with competing interactions, originally introduced in Ref.\citenum{PhysRevB.93.064430}
\begin{align}
\mc{F} = - \frac{J_1}{2} (\nabla \mb{m})^2 - \frac{J_2 a^2}{2} (\nabla^2 \mb{m})^2 - \frac{\mb{H}}{a^2} \cdot \mb{m} +\frac{K}{a^2} m_z^2 \,,
\label{eq:EnergyFunc}
\end{align}
where $J_1$, $J_2$, $H$ and $K$ are in units of [eV]. We introduce dimensionless variables $\mb{r} = \mb{\td{r}}/(\ell a)$, $t=\td{t}/\varepsilon_\Lambda$ and $\beta = \tilde{\beta} \varepsilon_\Lambda$, with $\ell$ and $\varepsilon_\Lambda$ a characteristic length and energy scale respectively. Stationary configurations of the action \eqref{eq:Action}, denoted as $\Phic$ and $\Pic$, are found by minimizing the energy functional, \textit{i.e} by solving the equations $\delta \mc{F}/\delta\Phic = 0 = \delta \mc{F}/\delta\Pic$. This class of solutions are characterized by a finite topological charge $Q = \frac{1}{4 \pi} \int d\mb{r} \mb{m} \cdot (\partial_x \mb{m} \times \partial_y \mb{m})$, 
and a magnetization profile that decays to zero at spatially infinity. Choosing $\ell = \sqrt{J_2/J_1}=1$, $\varepsilon_\Lambda = J_1$, and $\mb{H} = H \hat{z}$, the action of Eq.~\eqref{eq:Action} in reduced units is written as follows $\mc{S}=\bar{S} \int dt \int dr [\dot{\Phi}(\Pi-1) - \mc{F} (\Phi,\Pi)]$, with 
\begin{align}
\mc{F} = - (\nabla \mb{m})^2 /2- (\nabla^2 \mb{m})^2/2 -  h m_z + \tilde{\kappa} m_z^2 \,,
\label{eq:Energy}
\end{align}
with $\tilde{\kappa} = K /J_1$ and $h = H/J_1$. 

Rotationally symmetric solutions of the model \eqref{eq:Energy} are described by $\Phi(\mb{r})= -Q \phi$ and $\Theta(\mb{r}) = \Theta(\rho)$ with boundary conditions $\Thetac(0)=\pi$ and $\Thetac(\rho \rightarrow \infty) = 0$. Upon minimization of the energy $\int d\mb{r} \mc{F}$ we obtain a nonlinear Euler equation for the skyrmion profile $\Theta(\mb{r})$, which generally cannot be solved analytically. In the $\rho \gg 1$ limit we find the approximate solution $\Theta(\rho)= \Re[c_+ e^{-\gamma_+ \rho}+c_- e^{-\gamma_- \rho}]$, with $\gamma_{\pm} = \sqrt{-1\pm \tilde{\gamma}}/\sqrt{2}$ with $\tilde{\gamma}=\sqrt{1-4(h+2 \tilde{\kappa})}$ and $c_{\pm}$ constants. The skyrmion size is defined as $\lambda \equiv 1/\Re[\gamma_\pm]$. (see Ref.~\citenum{PhysRevB.93.064430} for more details on the model). For example for $K=0$ and $H=0.86$ T, the skyrmion size is $\lambda = 10a$. The rotationally symmetric skyrmion profile is depicted in Fig.~2-d). 

To generate the double-well potential term, essential for the construction of the helicity-qubit, we consider skyrmions characterized by elliptical profiles, which can be the result of defect engineering. The skyrmion profile is parametrized as $\Theta_{\ell} = \Theta(\rho) + g(\rho) \cos 2\phi$, where $g(\rho)$ is dictated by the microscopic mechanism responsible for skyrmion deformation. Here we use the phenomenological function $g(\rho)=\mbox{sech}[(\rho-\lambda)/\Delta_0]$, and the elliptical skyrmion is depicted in Fig.~3-d).

~\\
\textit{\textbf{Energy and Hamiltonian Terms}}. We summarize the energy terms on the classical level in physical units, the definition of dimensionless parameters and the corresponding quantized Hamiltonian terms. The various interactions of interest introduced in the main part of the manuscript are
\begin{align}
\mc{F}' = \bar{S} \int d{\mb{r}}&[ \frac{K}{a^2} m_z^2 - \frac{H}{a^2} m_z + \frac{K_x}{a^2} m_x^2 + \frac{H_{\perp}}{a^2} y m_x \nonumber \\
&- \mb{E} \cdot \mb{P} +\frac{B}{a^2} f(t) \cos(\omega t+\pext) x m_x] \,,
\end{align}
with $\mb{P}= [\hat{e}_x \times (\mb{m} \times \pt_x \mb{m})+ \hat{e}_y \times (\mb{m} \times \pt_y \mb{m})]$ the electric polarization, and $\mb{E} = E P_E a \hat{z}$ the electric field. $K$, $H$, and $K_x$ are in units of [eV], $H_{\perp}$ and $B$ in units of [eV/m], $P_E$ in units of [C/m$^2$], $E$ in units of V/m, and $a$ the lattice constant in units of [m]. Choosing $a$ as a typical length and $J_1$ as a typical energy scale, we arrive at
\begin{align}
\mc{F}' = \int d{\mb{r}}[& \tilde{\kappa} m_z^2 - h m_z + \tilde{\kappa}_x m_x^2 + \tilde{h}_\perp y m_x \nonumber \\
-& \varepsilon_z \hat{z} \cdot \mb{P} + \tilde{b} f(t) \cos(\omega t+\pext) x m_x] \,,
\end{align}
given now in dimensionless units $\tilde{\kappa}= \bar{S} K/J_1$, $\tilde{\kappa}_x= \bar{S} K_x/J_1$, $h=\bar{S} H/J_1$, $\tilde{h}_\perp=\bar{S}H_{\perp} a/J_1$, $\tilde{b}=\bar{S}B a/J_1$, and $\varepsilon_z = \bar{S} E P_E a^3/J_1$. Following the quantization procedure described in detail below, the quantum Hamiltonian in terms of $\vp$ and $S_z$ reads,
\begin{align}
\tilde{H} &= \kappa S_z^2 - h m_z + \kappa_x \cos 2 \vp - E_z \cos \vp \nonumber \\
&+ h_\perp \sin \vp + b f(t) \cos(\omega t+\pext)\,,
\end{align}
where $\kappa = \tilde{\kappa} \int_{\mb{r}} (1-\cos \Theta)^2/[\int_{\mb{r}} (1-\cos \Theta)]^2$, $\kappa_x  = (\tilde{\kappa}_x/4)\int_{\mb{r}} \sin 2\Theta g(\rho)$ , $E_z = \varepsilon_z \int_{\mb{r}} [\sin 2 \Theta/2\rho + \Theta']$, $h_\perp =(\tilde{h}_\perp/4) \int_{\mb{r}} \rho \sin \Theta$, and $b=( \tilde{b}/4)\int_{\mb{r}} \rho \sin \Theta$. 

\section{Skyrmion Quantization}
To investigate the quantum effects, we employ a functional integral formulation, in which the partition function is given by $Z=\int \mc{D} \mb{m}   e^{i \mc{S}(\mb{m},\mb{\dot{m}})}$. Here $\mc{S}=\int dt L$ is the action, with
\begin{align}
L=\bar{S} \int d\mb{r}[ \mc{A}(\mb{m}) \cdot \dot{\mb{m}} - N_A \mc{F} ]\,,
\end{align}
where $\mc{A}(\mb{m})=[1-\tilde{e}_\Phi \cdot (e_\Phi \times \mb{m})] e_\Phi /(\tilde{e}_\Phi \cdot \mb{m}))$ is the gauge potential. We use $\tilde{e}_\Phi=(\cos\Phi, \sin \Phi,0)$ and $e_\Phi=(-\sin\Phi, \cos \Phi,0)$, and we also note that $\mc{A} \cdot \dot{\mb{m}} = (1-\cos \Theta) \dot{\Phi}$. The commutation relations are $\{ m_i(\mb{r})  , m_j(\mb{r}') \} = \epsilon_{ijk} m_k(\mb{r}) \delta(\mb{r}-\mb{r}')$, where $\{A, B\}$ is the Poisson bracket satisfying
\begin{align}
\{ A(\mb{r}), B(\mb{r}') \} = \int d\mb{r}''[ \frac{\delta A(\mb{r})}{\delta \Phi(\mb{r}'')} \frac{\delta B(\mb{r}')}{\delta \Pi(\mb{r}'')} -\frac{\delta A(\mb{r})}{\delta \Pi(\mb{r}'')} \frac{\delta B \mb{r}'}{\delta \Phi(\mb{r}'')} ]
\end{align}  
provided that $\{ \Phi(\mb{r}),\Pi (\mb{r}') \} = \delta(\mb{r}-\mb{r}')$. The model $\mc{F}$ is characterized by an unbroken global symmetry, $\mb{m} \rightarrow \mc{M}(\vp(t)) \mb{m}$, with
\[
\mc{M} =
  \begin{bmatrix}
    \cos \vp & -\sin \vp  &0 \\
    \sin \vp & \cos \vp & 0\\
    0 &0 & 1 
  \end{bmatrix} \,.
\]

Instead of the original magnetization vector $\mb{m}$, it appears convenient to introduce $\mb{n} = \sqrt{1-\cos \Theta}/\sin \Theta~ \mb{m}$ and the corresponding gauge vector $\mc{A}_{\mb{n}} = \pt_{\Phi} \mb{n}$, such that the Wess Zumino term of the action remains unchanged, $\mc{A}_{\mb{n}} \cdot \dot{\mb{n}} = (1-\cos\Theta) \dot{\Phi}$. It is also important to note that the zero mode of the skyrmion associated with infinitesimal rotations is equal to $\mc{A}_{\mb{n}_0}= \pt_\Phi \mb{n}_0$, where $\mb{n}_0$ describes the skyrmion profile. In the naive perturbation expansion around the skyrmion, the zero mode leads to infrared divergences unless it is removed from the path integral by imposing proper constraints. 

Here we use a path integral quantization method according to which the collective coordinates are introduced by performing a canonical transformation of the dynamical variables in the phase space path integral\cite{PhysRevD.11.2943,PhysRevD.49.3598}. The zero mode is removed by introducing a $\delta$-function constraint of the form
\begin{align}
1 = \int \mc{D}\vp(t) J_{\vp} \delta (F_1) \,,
\label{eq:Const1}
\end{align}
with $F_1 = \int d\mb{r} \mc{A}_{\mb{n}_0} \cdot (\tmn -\tmn_0)$, $J_{\vp} = \delta F_1/\delta \vp$ is the Jacobian of the transformation, and we use the tilde notation for rotated vectors, $\tilde{B}= \mc{M} B$. The constraint ensures fluctuations of the magnetization field around the skyrmion are orthogonal to the rotational zero mode. We introduce an additional constraint related to the conservation of the conjugate to $\vp$ momentum,
\begin{align}
1 = \int \mc{D}S_z(t) J_{S_z} \delta (F_2) \,,
\label{eq:Const2}
\end{align}
with $F_2 = (1/\Lambda) \int d\mb{r} \mc{A}_{\mb{n}_0} \cdot (\tilde{\mc{A}}_{\mb{n}}-\tilde{\mc{A}}_{\mb{n_0}} )$, $J_{S_z} = \delta F_2/\delta S_z$, and $\Lambda = \int d\mb{r} \mc{A}_{\mb{n}_0} \cdot \mc{A}_{\mb{n}_0}$ a normalization constant. To ensure that the above change of variables constitutes a canonical transformation we introduce the following variables in the integration, $\mb{n} = \tilde{\mb{n}}_0+ \boldsymbol{\gamma}$ and $\mc{A}_{\mb{n}} = c \tilde{\mc{A}}_{\mb{n}_0} +  \boldsymbol{\zeta}$, with $c$ a constant to be specified from the momentum conservation constraint, $P - S_z = F_2 = 0$ with $P = \int d\mb{r} \mc{A}_{\mb{n}} \cdot \pt_{\vp} \mb{n}=\int d\mb{r} (1-\cos\Theta) \pt_{\vp}\Phi$ the total momentum. We note that it holds $\{ P, \Phi \} = -\pt_{\phi} \Phi$, confirming that $P$ plays the role of infinitesimal generator of rotations. After some straightforward calculation we find
\begin{align}
c = \frac{S_z -\int \boldsymbol{\zeta} \cdot \pt_{\vp} \mb{n}}{\int \mc{A}_{\mb{n}_0} \cdot \pt_{\Phi}\mb{n}}\,.
\end{align}
We confirm that the phase-space path integral retain its canonical form in terms of the new variables $\int_{\mb{r},t} \mc{A}_{\mb{n}} \cdot \dot{\mb{n}} =\int_t[ S_z \dot{\varphi_0} + \int_{\mb{r}} \boldsymbol{\zeta} \cdot \dot{\boldsymbol{\gamma}}]$, and also that the two Jacobian factors cancel $J_{\vp} J_{S_z}= 1$, with $J_{\vp} = \int \mc{A}_{\mb{n}_0} \cdot \pt_{\Phi_0} \mb{n}$. We note that $\boldsymbol{\zeta}$ and $\boldsymbol{\gamma}$ denote fluctuations around the gauge and magnetization vectors correspondingly, and can be associated to the fluctuations around fields $\Phi, \Theta$. 

The partition function is now written in terms of the new variables,
\begin{align}
Z= \int \mc{D} \vp \mc{D} S_z \mc{D} \boldsymbol{\zeta} \mc{D} \boldsymbol{\gamma} \delta(F_1) \delta(F_2)e^{i \mc{S}(\vp,S_z,\boldsymbol{\zeta}, \boldsymbol{\gamma} )}
\end{align}
with
\begin{align}
\mc{S} = \bar{S} \int_{t, \mb{r}}[S_z \dot{\vp} + \boldsymbol{\zeta} \cdot \dot{\boldsymbol{\gamma}}) - \mc{F}(\vp,S_z,\boldsymbol{\zeta}, \boldsymbol{\gamma} ) ] \,.
\end{align} 

Our current task is to analyze the energy functional $\mc{F}(\Pi,\Phi)$ of Eq.~\eqref{eq:EnergyFunc} in terms of the new variables $\vp, S_z, \boldsymbol{\zeta}$ and $\boldsymbol{\gamma}$. Since $P(t) = \int_{\mb{r}} \tilde{\Pi} \pt_{\phi} \Phi$, with $\tilde{\Pi}=1-\Pi$, we can apply the following transformation $\tilde{\Pi}(\mb{r},t) = [S_z(t)-\int_{\mb{r}} \eta(\mb{r},t) \pt_{\phi} \Phi] \tilde{\Pic}/\Lambda +\eta(\mb{r},t)$, with $\Lambda =\int_{\mb{r}} \tilde{\Pic} \pt_\phi \Phi$, while $\eta$ corresponds to quantum fluctuations around the classical configuration. Fluctuations around the field $\Phi$ are denoted as $\Phi = \Phic[\mb{r},\vp(t)] + \xi(\mb{r},t)$. It is easy to verify that with the above definitions the Wess-Zumino term maintains its canonical form $\int_{\mb{r},t} \tilde{\Pi} \dot{\Phi} =\int_{t} S_z\dot{\vp} + \int_{\mb{r},t} \eta \dot{\xi} $. To prove this relation we used the constraint $F_1$ of Eq.~\eqref{eq:Const1}, which in terms of the new variables takes the form $F_1 = \int (1-\Pic) \xi=0$. 

By implementing these changes of variables, the partition function is given as,
\begin{align}
Z= \int \mc{D} \vp \mc{D} S_z e^{i\int_t [\bar{S} S_z \dot{\vp} -\tilde{H}(\vp,S_z)]} \tilde{Z}[\vp, S_z] \,,
\label{eq:Partition}
\end{align}
where $\tilde{H} = \kappa S_z^2-h S_z -E_z \cos \Phi_0$ is the quantum Hamiltonian with $\kappa = \tilde{\kappa} \int_\mb{r} \tilde{\Pic}^2/\Lambda_0^2$, $\Lambda_0= \int \tilde{\Pic} $, and $E_z = \varepsilon_z \int d\mb{r}[\sin 2\Thetac /2 \rho +\Thetac']$. We note that we retain leading terms in powers of $1/\bar{S}$, and up to quadratic in $S_z$ and $\eta,\xi$. The fluctuating part of the partition function equals,
\begin{align}
\tilde{Z}= \int \mc{D} \chi \mc{D} \chi^{\dagger} \delta(F_1) \delta(F_2) e^{i\int_{\mb{r},t} \chi^{\dagger} [\mc{G} + \mc{K}] \chi}  \,,
\end{align}
with $\chi=(\eta,\xi)$. Here the operator $\mc{G}$ describes the magnon spectrum around the skyrmion, while $\mc{K}$ is responsible for the dynamical coupling of the skyrmion with the surrounding magnons. The fluctuating part of the partition function is written in a Gaussian form and can be manipulated within the real-time Keldysh functional integral. The analysis will generate a dissipative term as well as a Langevin random noise, which will play a role in the estimation of the skyrmion qubit decoherence time. These terms are omitted from the present analysis and are left for the future.  

Using standard equivalence between the path integral and canonical quantization, we introduce a collective coordinate operator $\hvp$ and its conjugate momentum $\hat{S}_z=(-i/\bar{S}) \pt_{\hvp}$ with $[\hvp, S_z] = i/\bar{S}$. If $E_z=0$, the momentum operator commutes with the corresponding Hamiltonian and stationary states are labeled by a conserved charge $s$ constrained to be an integer with $S_z \vert s \rangle = s/\bar{S} \vert s \rangle$. The phase space associated to $\hvp \vert \varphi_0 \rangle = \varphi_0 \vert \varphi_0 \rangle$ has a circular topology $\vert \varphi_0 \rangle= \vert \varphi_0 +2 \pi \rangle$. We also note that it holds $e^{\pm i \hvp} \vert s \rangle = \vert s\pm 1 \rangle$. 

\section{Basic Qubit Types}
\subsection{$S_z$-qubit}
In the presence of an out-of-plane uniform magnetic field, an easy-axis anisotropy and an out-of-plane electric field, the Hamiltonian for the helicity degrees of freedom reads,
\begin{align}
H_{S_z}=\kappa(\hat{S}_z -h/\kappa)^2 -E_z \cos \hvp \,.
\label{eq:ChargeH}
\end{align}
We note that Eq.~\eqref{eq:ChargeH} resembles the circuit Hamiltonian of the Cooper pair box, with $\kappa$ the charging energy, $h/\kappa$ the offset charge, and $E_z$ the Josephson energy. To determine the state $\vert s \rangle$ and eigenenergies $\mc{E}_s$, we solve the corresponding Schr{\"o}dinger equation for states $\Psi_s(\varphi_0) = \langle \varphi_0 \vert s \rangle$,
\begin{align}
\kappa[-\frac{i}{\bar{S}} \pt_{\vp} - h/\kappa]^2 \Psi_{s}(\varphi_0) - E_z\cos \vp \Psi_{s}(\varphi_0)  =\mc{E}_s \Psi_{s}(\varphi_0) \,,
\end{align}
with boundary conditions $\Psi_{s}(\varphi_0) = \Psi_{s}(\varphi_0+2\pi)$. For $\bar{h}\in [0,1/2]$, with $\bar{h} = \bar{S} h/\kappa$, functions $\Psi_s(\varphi_0)$ can be written using the Mathieu functions as 
\begin{align}
\Psi_s(\varphi) = e^{i \bar{h} \varphi_0} c_j \mc{M}_j  \left( \frac{4 \mc{E}_s }{\kappa},  \frac{2 E_z}{\kappa},\frac{\varphi_0}{2} \right) \,,
\end{align} 
where the index $j=C (S)$ represents the even (odd) solutions and $\mc{E}_s= (\kappa/4) \mc{M}_A [2(\kappa-\bar{h}),-2E_z/\kappa]$.

We now focus on the $E_z \ll \kappa$ and $\bar{h}= 1/2$ regime, where the lowest two levels are almost degenerate and seperated by a small energy internal controlled by the electric field. The skyrmion qubit states $\vert 0 \rangle$ and $\vert 1 \rangle$, with $\hat{S}_z \vert s \rangle =s/\bar{S} \vert s \rangle$, represent deviations of the $z$ component of the magnetization from equilibrium. For $E_z>0$, the degeneracy is lifted and the energy eigenstates are symmetric and antisymmetric superpositions of skyrmion qubit states,$(\vert 0\rangle \pm \vert 1\rangle)/\sqrt{2}$.

Truncating the full Hilbert space to the subspace spanned by these two states, one can write the qubit reduced Hamiltonian as 
\begin{align}
\hat{H}_q = \frac{H_0}{2} \hat{\sigma}_z - \frac{X_c}{2} \hat{\sigma}_x \,,
\label{eq:QubitHC}
\end{align}
with $H_0=\kappa(1-2 \bar{h})/\bar{S}$, and $X_c=E_z$ while the qubit energy levels are given by $\mc{E}_{\pm} = \pm \sqrt{H_0^2+X_c^2}/2$. 

\subsection{Helicity-qubit}

In this section we discuss how an elementary qubit described by a double-well potential profile can be constructed.  The helicity-qubit is a good example of how one can engineer the qubit properties through the choice of suitable parameters. To proceed, we consider a material with an easy-plane anisotropy, $\mc{F}_x = \bar{S} K_x/a^2 \int_{\mb{r}} m_x^2$, and a skyrmion with an elliptical profile, such as the one depicted in Fig.~3-d). The Hamiltonian reads $H_{\vp} = \kappa \hat{S}_z^2-h \hat{S}_z +V(\hvp)$, with the double-well potential given by 
\begin{align}
V(\hvp) =\kappa_x \cos 2\hvp -E_z \cos \hvp + h_\perp \sin \hvp \,. 
\end{align}

The elliptical deformation is parametrized as $\Thetac(\rho,\phi) = \Thetac(\rho) + g(\rho) \cos 2 \phi$, with $g(\rho)$ dictated by the microscopic mechanism responsible for deforming the skyrmion. Here we use $g(\rho)=\mbox{sech}[(\rho-\lambda)/\Delta_0]$, and $\kappa_x = (\bar{S} K_x/4 J_1) \int \sin 2\Thetac f(\rho)$. We note that a circular skyrmion profile with $g(\rho)$ fails to reshape potential landscape and produce the required $\cos 2\hvp$ term. A depth difference between the wells can be created by an in plane magnetic field gradient of the form $\mb{H}_{\perp}(\mb{r}) =H_\perp y \hat{x}$. We then find $h_\perp= -(\bar{S} H_\perp/4 J_1) \int \rho \sin\Thetac$. An application of a magnetic field gradient of the form $\mb{H}_{\perp}(\mb{r}) =H_\perp x \hat{y}$, results an asymmetric potential term of opposite sign $h_\perp =(H_\perp/4 J_1) \int \rho \sin\Thetac$. 

We seek eigenfunctions of $H_{\vp}$ as a linear combinations of the $2\pi$-periodic basis functions
\begin{align}
\Psi_n(\Phi_0) = \frac{1}{\sqrt{2\pi}} \sum_{m} c^n_m e^{i m \vp} \,,
\end{align}
which maintain the required $2\pi$-periodicity in $\vp$, and $n$ labels the $n$th eigenstate. The corresponding set of equations for coefficients $c^n_m$ reads,
\begin{align}
(\frac{\kappa}{\bar{S}^2} m^2- E_n - \frac{h}{\bar{S}} m) c ^n_m +\frac{\kappa_x}{2}(c ^n_{m+2}+c ^n_{m-2}) \nonumber \\
-\frac{E_z}{2} (c ^n_{m+1}+c ^n_{m-1})+\frac{h_\perp}{2} (c ^n_{m+1}-c ^n_{m-1})=0\,.
\end{align} 

A numerical diagonalization keeping up to $\vert m\vert= 50$  terms yields the eigenenergies and the corresponding coefficients $c^n_m$. In Fig.~3-b) we plot the lowest three eigenvalues together with the potential $V(\vp)$, with a minimum at $\varphi_m= \tan^{-1}(\sqrt{16 \kappa_x^2-E_z^2}/E_z)$. All three $E_z$, $h_y$, and $h$ are external parameters to adjust the energy levels. For $E_z=0$ there is a degeneracy point at  $\bar{h}=\bar{S} h/\kappa =1$. For $h_\perp=0$, the two lowest energy functions $\Psi_{0,1}$ are symmetric and antisymmetric combinations of the two wavefunctions localized in each well, while for $h_\perp >0$, they are localized in different wells.

In the limit $\bar{h}-1 \ll 1$, we can reduce the analysis to the two lowest levels and derive the  qubit Hamiltonian,
\begin{align}
\hat{H}'_q= \frac{H_0}{2}\hat{\sigma}_z -\frac{X_c }{2}\hat{\sigma}_x  \,,
\label{eq:QubitHF}
\end{align} 
provided that $H_0 =(\mc{E}_1-\mc{E}_0)$ and $X_c = g_e E_z$ for $h_\perp=0$, or  $X_c = g_b h_\perp$ for $Ez=0$. Constants $H_0$, $g_e$ and $g_b$ are found numerically. 

\section{Qubit Control}

In this section we discuss how skyrmion qubits can be manipulated to implement quantum algorithms, with the techniques introduced here being applicable to both skyrmion types. The predominant protocol is via microwave magnetic field gradients with frequencies at the qubit transition $\omega_q$. A magnetic field gradient couples with the magnetization as $\Fext= \int d\mb{r} \mb{B}(\mb{r},t) \cdot \mb{m}(\mb{r},t)$, with $\mb{B}(\mb{r},t)=B/a^2  \cos(\omega t+\phi)f(t) x \hat{e}_x$. Additional Hamiltonian terms appear, $\Hext(t) = b \cos (\omega t+\phi) f(t)\cos \vp$, with $b = (B/4 J_1) \int_{\mb{r}} \rho \sin \Thetac$. In terms of the reduced quantum Hamiltonian we find,
\begin{align}
\Hext^q = b_x(t) \hat{\sigma}_x  \,,
\end{align}
where  $b_x(t)= b_0 \cos (\omega t+\phi)f(t)$ and $b_0= b \Re(\langle 1 \vert \cos \vp \vert 0 \rangle +\langle 0 \vert \cos \vp \vert 1\rangle)/2$. Here $\vert 0 \rangle$ and $\vert 1 \rangle$ denote the lowest two qubit states. 

The eigenvectors of the unperturbed qubit Hamiltonian $H_q$ of Eq.~\eqref{eq:QubitHC} are
\begin{align}
\vert \Psi_{-} \rangle &= \cos \frac{\theta}{2} \vert 0 \rangle + \sin \frac{\theta}{2} \vert 1 \rangle \nonumber \\
\vert \Psi_{+} \rangle &= \sin \frac{\theta}{2} \vert 0 \rangle - \cos \frac{\theta}{2} \vert 1 \rangle  \,,
\label{eq:Eigenstate}
\end{align}

with $\tan \theta = X_c/H_0$, while the corresponding eigenvalues are $\mc{E}_{\pm}=\pm \sqrt{H_0^2+X_c^2}/2$ and energy splitting $\omega_q = (\mc{E}_+ - \mc{E}_-)$. In the $\vert \Psi_{\pm} \rangle$ basis, the driven qubit Hamiltonian takes the form
\begin{align}
H_q = \frac{\omega_q}{2} \hat{\sigma}_z + b_x(t)[ \cos \theta \hat{\sigma}_x +\sin \theta \hat{\sigma}_z] \,.
\end{align}

We note that close to the degeneracy point $X_c \ll H_0$ (so called "sweet spot"), it holds $\sin \theta \ll \cos \theta$. To elucidate the role of the drive, we transform $H_q$ into the rotating frame at a frequency $\omega$,
\begin{align}
\Hrot = \frac{\Delta \omega}{2} \hat{\sigma}_z +\frac{\Omega f(t)}{2}[\cos \phi \hat{\sigma}_x + \sin \phi \hat{\sigma}_y] \,,
\label{eq:HRot}
\end{align} 
with $\Delta \omega = \omega_q - \omega$ the detuning frequency and $\Omega = b_0 \cos\theta$. When $f(t)=1$, the eigenvectors of the rotated Hamiltonian are of the form 
\begin{align}
\vert \tilde{\Psi}_{-} \rangle &= \cos \frac{\tilde{\theta}}{2} \vert \Psi_- \rangle + \sin \frac{\tilde{\theta}}{2} \vert \Psi_+ \rangle \nonumber \\
\vert \tilde{\Psi}_{+} \rangle &= \sin \frac{\tilde{\theta}}{2} \vert \Psi_- \rangle - \cos \frac{\tilde{\theta}}{2} \vert \Psi_+ \rangle  \,,
\label{eq:EigenstateR}
\end{align}
where $\tan \tilde{\theta} = \Omega/\Delta \omega$. The corresponding energy levels are $\mc{\tilde{E}}_{\pm} = \pm \sqrt{\Delta \omega ^2 +\Omega^2}/2$. If at initial time we start from the state $\vert \tilde{\Psi}_- \rangle$, the probability to find the system at the state $\vert \tilde{\Psi}_+ \rangle$ is given by $P(t) = \Omega^2/\tilde{\Omega}^2 \sin^2(\tilde{\Omega} t/2)$, with $\tilde{\Omega}= \sqrt{\Delta \omega ^2 +\Omega^2}$ the Rabi frequency. Single-qubit operations correspond to rotations of the qubit state by a certain angle about a particular axis, as the result of a unitary operator applied to the target qubit. As an example, for \textit{in-phase} pulses $\pext=0$, and resonant driving on the qubit energy splitting, $\Delta \omega=0$, the unitary operator $U_x(t) = e^{-\frac{i}{2}\vartheta(t) \hat{\sigma}_x}$ corresponds to rotations around the x-axis by an angle $\vartheta(t)= -\Omega \int_0^{t} f(t') dt'$ \cite{doi:10.1063/1.5089550}. \textit{Out-of-phase} pulses $\pext =\pi/2$ correspond to rotations of the qubit state about the $y$ axis. 

\section{Relaxation Mechanisms}

The scope of this section is to calculate the relaxation and decoherence rates, which as we demonstrate, are directly proportional to the spectral densities of the random noises acting on the qubit. To include classical noise sources we consider the magnetization dynamics encoded in the Landau-Lifshitz-Gilbert equation (LLG), $\dot{\mb{m}}= \gamma \mb{\Feff} \times \mb{m} +\alpha \mb{m} \times \mb{\dot{m}}$, where $ \mb{\Feff}= -\delta \mc{F}/\delta\mb{m}$, and $\alpha$ is the Gilbert damping constant. Starting from the LLG equation, it is possible to derive the equation of motion for the collective coordinates $\zeta= \vp, S_z$\cite{PhysRevLett.100.127204},
\begin{align}
G_{ij} \dot{\zeta}_j + F_i - \alpha_{i} \dot{\zeta}_i =0  \,,
\end{align}

where $G_{ij}=1=-G_{ji}$ is the gyrotropic tensor, $F_i = \pt \mc{F}/\pt \zeta_i$ is the generalized force, and $\alpha_i$ is the damping tensor given by $\alpha_{i}= \alpha \bar{S} \int \pt \mb{m}/\pt \zeta_i \cdot \pt \mb{m}/\pt \zeta_i$. In particular we find, $\alpha_{\vp} =\bar{S} \alpha \int_{\mb{r}} \sin \Theta$ and $\alpha_{S_z} =\bar{S} \alpha \int_{\mb{r}}(1-\cos \Theta)^2/\Lambda_0^2$, with $\Lambda_0 = \int_{\mb{r}}(1-\cos \Theta)$. Thus, the motion of both the helicity $\vp$ and the conjugate momentum $S_z$ is governed by Ohmic dissipation terms with corresponding dissipation constants $\alpha_{\vp}$ and $\alpha_{S_z}$ respectively. They are accompanied by random fluctuating forces, which enter the quantum Hamiltonian as $\hat{H} \rightarrow \hat{H} + \xi_{\vp} \vp + \xi_{S_z} \hat{S}_z$, with $\xi_i$ fully characterized by the classical ensemble averages $\langle \xi_i(t) \rangle=0$ and $\langle \xi_i(t) \xi_j(t') \rangle= \delta_{ij}S_i(t-t')$. The dissipative kernel $\alpha_{i}$ and the correlator $S_i(t-t')$ are related via the fluctuation-dissipation theorem, 
\begin{align}
S_i(\omega) = \alpha_{i} \omega \coth(\frac{\beta \omega}{2})\,,
\end{align}
with $S_i(t) =\int d\omega/2\pi e^{-i \omega t} S_i(\omega)$. 

We now seek the reduced form the fluctuating fields enter the qubit Hamiltonian, a procedure that is identical for both skyrmion qubit categories. In the subspace spanned by states $\vert 0 \rangle$ and $\vert 1 \rangle$ we find
\begin{align}
H_q = \frac{H_0}{2} \hat{\sigma}_z-  \frac{X_c}{2} \hat{\sigma}_x + \xi_{\zeta}(t) \gamma_{\zeta,i} \hat{\sigma}_i  \,,
\end{align}
with $i=x,y,z$ and $\zeta = \vp, S_z$. Constants $\gamma_{\zeta,i}$ are $\gamma_{\zeta,x}= \Re[\langle 1 \vert \hat{\zeta} \vert 0 \rangle +\langle 0\vert \hat{\zeta} \vert 1 \rangle]/2$, $\gamma_{\zeta,y}= \Im[\langle 1 \vert \hat{\zeta} \vert 0 \rangle -\langle 0\vert \hat{\zeta} \vert 1 \rangle]/2$, and $\gamma_{\zeta,z}= \Re[\langle 1 \vert \hat{\zeta} \vert 1 \rangle -\langle 0\vert \hat{\zeta} \vert 0 \rangle]/2$. 

Typically, the dynamics of two-level systems are expressed in terms of two rates: the longitudinal relaxation rate $\Gamma_1=T_1^{-1}$ and the dephasing rate $\Gamma_2=T_2^{-1}$. The latter is a combination of effects of the depolarization $\Gamma_1$ and of the pure dephasing $\Gamma_\varphi$, combined to a rate $\Gamma_2= \Gamma_1/2+\Gamma_\varphi$. According to the Bloch-Redfield theory it holds, 
\begin{align}
\Gamma_1&= [\alpha_{\vp} (\gamma^{\perp}_{\vp})^2 +\alpha_{S_z} (\gamma^{\perp}_{S_z})^2 ] \omega_q \coth(\frac{\beta \omega_q}{2})
\nonumber \\
\Gamma_\varphi &= [\alpha_{\vp} (\gamma^{\parallel}_{\vp})^2 +\alpha_{S_z} (\gamma^{\parallel}_{S_z})^2 ] 2/\beta\,,
\end{align}
provided that $\gamma^{\perp}_{\zeta}=\cos \theta \gamma_{\zeta,x}+\gamma_{\zeta,y}+\cos \theta \gamma_{\zeta,z}$ and $\gamma^{\parallel}_{\zeta}=\cos \theta \gamma_{\zeta,z}+\sin \theta \gamma_{\zeta,x}$. An estimate of $T_1$ and $T_2$ in physical units is given in Table II of the main manuscript. 

\end{document}